\documentclass[twocolumn,prb,aps,showpacs]{revtex4-1}
\usepackage{textcomp,amssymb,graphicx}
\usepackage{hyperref}
\usepackage{amsbsy}
\usepackage{amsmath}
\usepackage{amssymb}
\usepackage[dvips]{color}
\usepackage{hyperref}
\usepackage{float}
\begin{document}
\title{Study of long-range orders of hard-core bosons coupled to
 cooperative normal modes  in two-dimensional lattices}
\author{A. Ghosh}
\author{S. Yarlagadda}
\affiliation{
CMP Div., Saha Institute of Nuclear Physics, HBNI,
Kolkata, India}

\begin{abstract}

{
{Understanding the microscopic mechanism of coexisting long-range orders
(such as lattice supersolidity)
in strongly correlated systems is a subject of immense  interest.
We study the possible manifestations of long-range orders, including lattice-supersolid phases with differently
broken symmetry,  in a two-dimensional square lattice system of hard-core bosons (HCBs) coupled
to archetypal cooperative/coherent normal-mode distortions such as those in perovskites.
At strong HCB-phonon coupling,
using a duality transformation to map the strong-coupling problem to a weak-coupling one,
we obtain an effective Hamiltonian involving nearest-neighbor, next-nearest-neighbor,
and next-to-next-nearest-neighbor hoppings and repulsions. Using stochastic series expansion
quantum Monte Carlo, we construct the phase diagram of the system.
As coupling strength is increased,
we find that  the system
undergoes a 
{first-order} quantum phase transition from a superfluid to a checkerboard solid 
at half filling and from a superfluid to a diagonal striped solid [with crystalline ordering wavevector
$\vec{Q}=(2\pi/3,2\pi/3)$ or $(2\pi/3,4\pi/3)$] at one-third filling
without showing any evidence of supersolidity.
On tuning the system away from these commensurate fillings, 
 checkerboard supersolid
is generated
near half filling whereas a rare diagonal striped supersolid
is realized  near one-third filling. Interestingly, there is an asymmetry in the
extent of supersolidity about one-third filling.
Within our framework, we also provide an explanation
for the observed checkerboard and stripe formations
in ${\rm La}_{2-x}{\rm Sr}_x{\rm NiO_4}$ at $x=1/2$ and $x=1/3$.}
}
\end{abstract}
\maketitle
\section{Introduction}
{The origin and character of lattice supersolidity \cite{lattice_supersolidity}
[i.e., the single-phase
 coexistence of superconductivity/superfluidity 
and charge density wave (CDW) realized in discrete lattices]
in naturally formed 
and artificially designed systems 
is  a central issue in condensed matter physics.
While phenomenological pictures \cite{bilbro,bismuthate3} exist to explain lattice-supersolidity,
a microscopic  theory that elucidates the homogeneous coexistence
is yet to be formulated.
Supersolidity is observed in a variety of lattice systems
such as the three-dimensional doped ${\rm BaBiO_3}$ \cite{bismuthate3,bismuthate2}; 
the layered dichalcogenides\cite{quasi_2d_1} and molecular crystals\cite{quasi_2d_2}; and the
quasi-one-dimensional doped trichalcogenide ${\rm NbSe_3}$\cite{quasi_1d_1} and 
doped spin ladder ${\rm Sr_{14}Cu_{24}O_{41}}$\cite{quasi_1d_2,quasi_1d_3}. Of importance
are the class of
materials that display superconductivity and diagonal long-range order
due to strong  electron-phonon interaction such as ${\rm K}$ or ${\rm Pb}$ doped ${\rm BaBiO_3}$
(where a  $10\%$ change in the ${\rm Bi-O}$ bond length\cite{bismuthate1} has been observed) 
and the alkali metal fullerides \cite{fullerenes}.
Interestingly, ${\rm BaBiO_3}$ assumes perovskite structure with two adjacent oxygen octahedra sharing
an oxygen leading to a cooperative breathing mode (CBM). Furthermore, ${\rm BaBiO_3}$ displays
valence disproportionation
with local cooper pairs [i.e., hard-core bosons (HCBs)] being formed and these HCBs
couple to the CBM
\cite{varma}.}

{As regards artificially engineered systems,
cold bosonic atoms in optical lattices provide a fertile playground for actualizing
exotic phases such as  lattice-supersolid phases with differently broken symmetry. In fact, only recently
 supersolidity was experimentally produced in an optical lattice 
 by generating effective long-range interactions using a vacuum mode of an optical cavity \cite{expt_supersolid}.
 On the theoretical side, lattice supersolidity has been realized  in two-dimensional (2D) square 
 \cite{scalettar1, scalettar2, boninsegni1, zoller, wessel1, pinaki, troyer1, 
  kar,lv2},
 triangular \cite{wessel2, kedar1, kedar2, melko1, boninsegni2, melko2, troyer2}
  and honeycomb \cite{wessel3, ye} lattices as well as in 
 a one-dimensional lattice \cite{mishra, ghosh, nigel}.
 By using extended boson Hubbard models involving hard-core bosons,
 while a supersolid has been been produced at a commensurate filling (i.e., half filling)
 in frustrated systems such as triangular lattices, commensurate supersolid has been unobtainable
 in unfrustrated systems such as square lattices. 
 On the other hand, supersolids can be realized in square lattices at incommensurate fillings
 by a mechanism
where 
{particles (i.e., interstitials) or holes (i.e., vacancies)} doped into a perfect crystal 
form a condensate by delocalizing in the
crystalline order.
Furthermore,  although
 striped supersolidity has been achieved in Refs. \onlinecite{boninsegni1,wessel1} on square lattices, it is 
nondiagonal and characterized by density ordering wavevector  $(\pi,0)$ or $(0,\pi)$.
  Even though diagonal stripes [characterized by
 crystalline ordering wavevector 
 $(2\pi/3,2\pi/3)$ or $(2\pi/3,4\pi/3)$]
 have been observed in systems such as ${\rm La}_{2-x}{\rm Sr}_x{\rm NiO_4}$ (LSNO) at $x=1/3$ hole doping 
 \cite{stripe_review, cheong0, cheong, tokura2, tokura3, tokura4,freeman} and predicted theoretically for long-range interactions in a lattice gas model
 at one-third
 filling \cite{zaanen},
 so far the corresponding  diagonal striped supersolid (dsSS) has been elusive on a square lattice
 (that is not subject to an external potential).
 Additionally, whether a cooperative electron-phonon interaction (that involves
 cooperative Jahn-Teller distortions) can explain the observed stripe charge order in LSNO is a controversial issue
 \cite{dagotto,hatsugai,oles}. }
 
{In the class of extended boson Hubbard models 
 of the type $t_1-t_2-...-t_m-V_1-V_2-...-V_n$ [involving hoppings $t_1$, $t_2$, $t_3$, etc. and interactions
 $V_1$, $V_2$, $V_3$, etc. of  ranges 
 nearest neighbor (NN), next-nearest neighbor (NNN), next-to-next-nearest neighbor (NNNN), etc.]
 on a square lattice, the minimum model for realizing a checkerboard supersolid (cSS) is the
 $t_2-V_1$ model \cite{unpub,min_cb}.
  It has also been shown that 
{  star/stripe  supersolid} [corresponding to crystalline ordering wavevector
 $(\pi,0)$ or $(0,\pi)$]
 can be realized in a $t_1-V_1-V_2$ model;  at one-fourth filling,
 a star solid results which is asymmetric with respect 
 to doping with interstitials and vacancies \cite{boninsegni1}.
  Identifying the relevant extended boson Hubbard model for obtaining the dsSS
   around  one-third filling and characterizing the state are still  open problems.}

{Here, inspired by the doped bismuthate systems, we develop
a microscopic theory of HCBs strongly coupled to the cooperative
breathing mode in a 
2D perovskite lattice. The  effective Hamiltonian for the HCBs is shown to be
an extended boson Hubbard model 
 of the form $t_1-t_2-t_3-V_1-V_2-V_3$. The $V_1$, $V_2$, and $V_3$ repulsive interactions correspond to
 the  minimum interactions needed to realize  the diagonal striped-order at one-third filling.  
 Unlike many lattice
models of the extended boson Hubbard type, the parameters (i.e., hopping term, strength of HCB-phonon coupling,
and phonon frequency) in our $t_1-t_2-t_3-V_1-V_2-V_3$ model either can  be determined
from band-structure calculations or can be obtained
from experiments.
 Supersolidity in our model
  results only away from one-third filling and is shown to be asymmetric with respect to
 doping the commensurate diagonal-striped solid (dsS) with vacancies and interstitials.
  Although checkerboard supersolidity (away from half-filling) and diagonal striped supersolidity
  (away from one-third filling)
  are realized, there is no direct supersolid-supersolid phase transition between the two  phases.
 We also show that our cooperative HCB-phonon framework can be extended to 
 study  charge order in LSNO; we demonstrate that
 the observed diagonal-stripe order at one-third filling and the checkerboard order
 at half filling in LSNO
 can be explained by invoking cooperative Jahn-Teller effect.}

The paper is organized as follows. In Sec. \ref{Eff_Hamiltonian}, we derive 
an effective Hamiltonian of the system using a non-perturbative treatment.
Next, in Sec. \ref{Numerical_calc} we briefly describe the numerical procedure, as well as the quantities/parameters
used in our study. Then, we discuss the results in Sec. \ref{results}, followed by a 
comparison with experimental observations in Sec. \ref{comparison}. 
Finally, in Sec. \ref{Conclusion}, we conclude.

\section{Effective Hamiltonian}\label{Eff_Hamiltonian}
\begin{figure}[t]
\includegraphics[width=1\linewidth,angle=0]{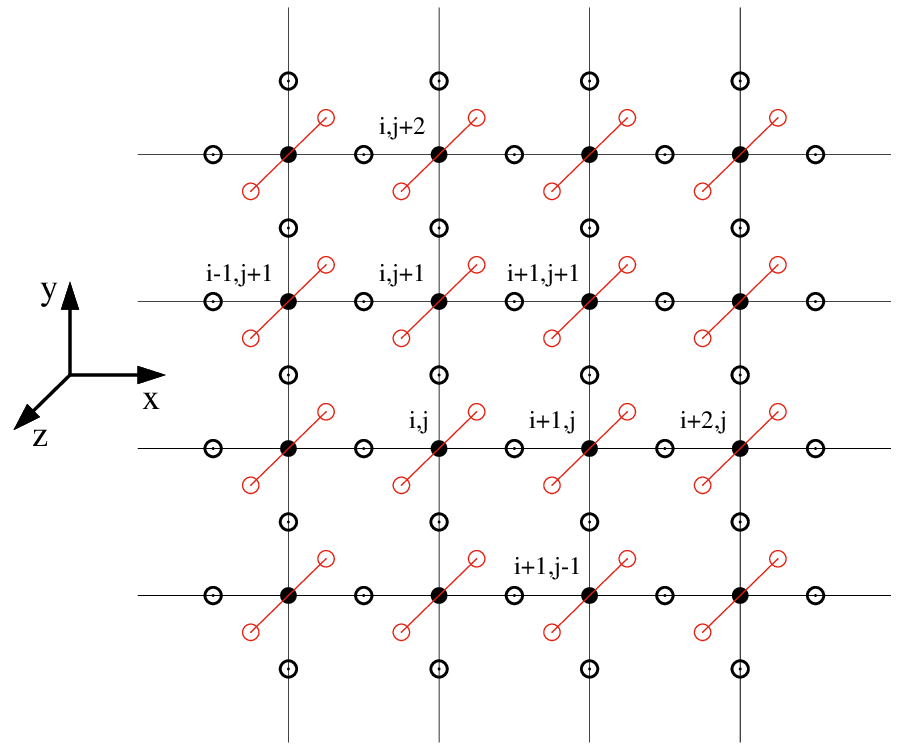}
\caption{(Color online) Two-dimensional cooperative breathing mode (CBM) system with hopping sites of hard-core-bosons (filled circles), in-plane oxygen atoms (black empty circles) and out-of-plane oxygen atoms (red empty circle). Only the in-plane oxygens 
{are involved in cooperative distortions}.}
\label{2d_CBM_fig}
\end{figure}
We start with a 
2D model of HCBs depicted in Fig. \ref{2d_CBM_fig}. The HCBs interact with the in-plane (xy) oxygen atoms via CBM, whereas the nature of the interaction is non-cooperative in the case of the out-of-plane oxygen atoms in the z-direction. The Hamiltonian of such a system can be written as $H=H_t+H_I+H_l$, where the hopping term $H_t$ is given by
\begin{align}
 H_t=-t\sum_{i,j}\left(d_{i+1,j}^\dagger d_{i,j}+d_{i,j+1}^\dagger d_{i,j}+{\rm H.c.}\right),
\end{align}
with $d_{i,j}(d_{i,j}^\dagger)$ being the destruction (creation) operator of a HCB at the hopping site $(i,j)$
{and $t$ being the hopping integral}.
The second term $H_I$ in the Hamiltonian, which represents the HCB-phonon interaction, has the form
\begin{align}
\!\!\!\!\!\!\!\!
H&_I=-g\omega_0\sum_{i,j}\Big[(a_{x;i,j}^\dagger+a_{x;i,j})(n_{i,j}-n_{i+1,j})\nonumber\\
+&(b_{y;i,j}^\dagger+b_{y;i,j})(n_{i,j}-n_{i,j+1})+\gamma(c_{z;i,j}^\dagger+c_{z;i,j})n_{i,j}\Big],\label{hcb-ph}
\end{align}
where $\gamma=\sqrt{2}$,
{$g$ is the HCB-phonon coupling constant, and $\omega_0$ is the optical-phonon frequency}. The terms $(a_{x;i,j}^\dagger+a_{x;i,j})/\sqrt{2M\omega_0}$ and $(b_{y;i,j}^\dagger+b_{y;i,j})/\sqrt{2M\omega_0}$ denote the displacement of the oxygen atom that is next to the $(i,j)$-th hopping site and in the positive x- and y-directions, respectively; here, $M$ is the mass of oxygen atom. The relative displacement of the two out-of-plane oxygens next to the $(i,j)$-th site couples to the HCB at $(i,j)$-th site and is denoted by $(c_{z;i,j}^\dagger+c_{z;i,j})/\sqrt{2\frac{M}{2}\omega_0}$ with $M/2$ being the reduced mass of the oxygen pairs. The expressions $(n_{i,j}-n_{i+1,j})$ and $(n_{i,j}-n_{i,j+1})$ in the first and second terms of Eq. (\ref{hcb-ph}) take care of the cooperative HCB-phonon interaction along the x- and y-directions, respectively. In the third term, note that we have only $n_{i,j}$ because of the non-cooperative nature of the HCB-phonon interaction along the z-direction. 
Furthermore, the last term in the Hamiltonian (i.e., the lattice term $H_l$), representing 
simple 
harmonic oscillators, is of the 
form
\begin{align}
 H_l=\omega_0\sum_{i,j}\left(a_{x;i,j}^\dagger a_{x;i,j}+b_{y;i,j}^\dagger b_{y;i,j}+\eta c_{z;i,j}^\dagger c_{z;i,j}\right),
\end{align}
with $\eta=1$.

{We consider systems in the  non-adiabatic regime ($t/\omega_0 \leqslant 1$) and strong-coupling region (large $g^2$).
To produce an effective polaronic Hamiltonian, 
we employ a duality transformation where the strong-coupling problem in the original frame
of reference [with small parameter $\propto (g\omega_0)/t$] is transformed into a weak-coupling 
problem in a dual frame of reference [with small parameter $\propto t/(g\omega_0)$, i.e.,
inverse of the small parameter in the original frame of reference].
To achieve the above end,}
we need to modify the 
Lang-Firsov transformation\cite{Lang_Firsov} so as to take into account the cooperative 
nature of the distortions along the x- and y-directions and non-cooperative nature in the z-direction. 
This involves the following canonical transformation $\tilde{H}=\exp(S) H \exp(-S)$ where $S$ is given by
\begin{align}
 S&=-g\sum_{i,j}\Big[(a_{x;i,j}^\dagger-a_{x;i,j})(n_{i,j}-n_{i+1,j})\nonumber\\
 &+(b_{y;i,j}^\dagger-b_{y;i,j})(n_{i,j}-n_{i,j+1})+\gamma(c_{z;i,j}^\dagger-c_{z;i,j})n_{i,j}\Big].
\end{align}
The transformed Hamiltonian can be written as $\tilde{H}=H_0+H_1$, where the unperturbed Hamiltonian is given by
\begin{align}
 H_0&=\omega_0\sum_{i,j}\left(a_{x;i,j}^\dagger a_{x;i,j}+b_{y;i,j}^\dagger b_{y;i,j}+\eta c_{z;i,j}^\dagger c_{z;i,j}\right)\nonumber\\
 &-E_p\sum_{i,j}n_{i,j}+2V_p\sum_{i,j}\left(n_{i,j}n_{i+1,j}+n_{i,j}n_{i,j+1}\right)\nonumber\\
 &-te^{-(E_p+V_p)/\omega_0}\sum_{i,j}\left(d_{i+1,j}^{\dagger}d_{i,j}+d_{i,j+1}^{\dagger}d_{i,j}+{\rm H.c.}\right),\label{unperturbed}
\end{align}
and the perturbation by
{
\begin{align}
 H_1=\sum_{i,j}H_{1i,j} ~~~~~~~~~&\nonumber \\
 =  -te^{-(E_p+V_p)/\omega_0}&\sum_{i,j}\Big[d_{i+1,j}^{\dagger}d_{i,j}\left({\tau^{ij}_{+x}}^{\dagger}\tau^{ij}_{-x}-1\right)\nonumber\\
 &+d_{i,j+1}^{\dagger}d_{i,j}\left({\tau^{ij}_{+y}}^{\dagger}\tau^{ij}_{-y}-1\right)+{\rm H.c.}\Big],
  \label{perturbed}
 \end{align}
}
where
\begin{align}
\tau^{ij}_{\pm x}&=\exp \Big[\pm g(2a_{i,j}-a_{i-1,j}-a_{i+1,j})\nonumber\\
&\pm g(b_{i+1,j-1}+b_{i,j}-b_{i,j-1}-b_{i+1,j})\pm \gamma g(c_{i,j}-c_{i+1,j})\Big],\nonumber
\end{align}
and
\begin{align}
\tau^{ij}_{\pm y}&= \exp \Big[\pm g(2b_{i,j}-b_{i,j-1}-b_{i,j+1})\nonumber\\
&\pm g(a_{i-1,j+1}+a_{i,j}-a_{i-1,j}-a_{i,j+1})\pm \gamma g(c_{i,j}-c_{i,j+1})\Big].\nonumber
\end{align}
Here $E_p=(4+\gamma^2)g^2\omega_0$ is the polaronic energy and $2V_p=2g^2\omega_0$ represents the nearest-neighbor repulsion for the HCBs.

The eigenstates of the unperturbed Hamiltonian $H_0$, relevant for perturbation theory are 
$|n,m\rangle=|n\rangle_{ hcb}\otimes |m\rangle_{ph}$, with $|0,0\rangle$ being the ground state with no phonons. The corresponding eigenenergies of such states are given by $E_{n,m}=E_n^{ hcb}+E_m^{ph}$. Similar to the case of 
one-dimensional CBM model \cite{rpsy}, we also have 
{$\langle n,0|H_1|n,0\rangle=0$,  which yields the first-order perturbation term $\langle 0,0|H_1|0,0\rangle=0$. 
In the region of interest in the parameter space, we note that
$te^{-(E_p+V_p)/\omega_0}<<\omega_0$; we perform second order perturbation theory similar to that in 
the 1D CBM model \cite{rpsy} and obtain the effective Hamiltonian to be
\begin{align}
 H_{\rm{eff}}=\langle 0|_{ph} H_0 |0\rangle_{ph} + H^{(2)} , \label{H_eff}
 \end{align}
 where
 \begin{eqnarray}
  H^{(2)} =
 \sum_{i,j,k,l}\sum_m\frac{\langle 0|_{ph} H_{1i,j} |m\rangle_{ph}\langle m|_{ph} H_{1k,l} |0\rangle_{ph}}{E_0^{ph}-E_m^{ph}}.
 \label{H2}
 \end{eqnarray}
One can easily see that the first term in $H_{\rm{eff}}$ is
\begin{align}
&\langle 0|_{ph} H_0 |0\rangle_{ph}=-E_p\sum_{i,j}n_{i,j}\nonumber\\
&+2V_p\sum_{i,j}(n_{i,j}n_{i+1,j}+n_{i,j}n_{i,j+1})\nonumber\\
&-te^{-(E_p+V_p)/\omega_0}\sum_{i,j}\left(d_{i+1,j}^{\dagger}d_{i,j}+d_{i,j+1}^{\dagger}d_{i,j}+{\rm H.c.}\right),
\end{align}
whereas the simplification of the second term (i.e, $H^{(2)}$) requires quite a bit of algebra.
We extend the derivation of the effective Hamiltonian for the 1D CBM case \cite{rpsy} to our 2D
case as well. 
As shown by using Schrieffer-Wolff transformation in
Appendix A of Refs. \onlinecite{sr1,sr2},  since $te^{-(E_p+V_p)/\omega_0}<<\omega_0$,
$H_{\rm{eff}} $ represents the exact 
Hamiltonian up to second order in perturbation.}
The small parameter here is given by $\left[\frac{t^2}{2(E_p+
V_p)\omega_0}\right]^{\frac{1}{2}}$ whose derivation is similar to that in Ref. \onlinecite{amit_ys}. For the second term $H^{(2)}$ in $H_{\rm{eff}}$, we  obtain the terms given in the following subsections.

\subsection{Nearest-neighbor (NN) repulsion}\label{subsec_NN_repul}
The NN repulsion term comes from a process where a particle jumps to a neighboring site and comes back. 
In 2D, this term further consists of two parts: $\sum\limits_{i,j}\left[n_{i,j}(1-n_{i+1,j})+n_{i+1,j}(1-n_{i,j})\right]$ and $\sum\limits_{i,j}\left[n_{i,j}(1-n_{i,j+1})+n_{i,j+1}(1-n_{i,j})\right]$. 
{Following a procedure explained in Appendix A, we get the expression for this process to be 
\begin{align}
-V_z\sum\limits_{i,j}[n_{i,j}(1-n_{i+1,j})+n_{i,j}(1-n_{i,j+1})] , \label{NN_repulsion}
\end{align}
with 
$V_z\approx \frac{2t^2}{2E_p+2V_p}$. 
The denominator $2E_p+2V_p$ in
$V_z$ is the difference of the energy of the intermediate state (i.e., $E_p+2V_p$
corresponding to the particle in the intermediate site) and the energy
of the initial state ($-E_p$). The exact expression for $V_z$ is derived
in Appendix A.
}

\subsection{Next-nearest-neighbor (NNN) and next-to-next-nearest-neighbor (NNNN) repulsions}\label{subsec_NNN_repul}
We first make an important point while considering a process of a particle hopping to a neighboring site and coming back. In 2D, excluding the originating site, we must take into account the occupancy information about all the three remaining NN sites of the intermediate site of 
{the hopping  process}. For example, consider a process where a HCB at site $(i,j)$ hops to its neighboring site $(i+1,j)$ and comes back. For this process, we need to keep in mind the occupancy of the sites $(i+2,j)$, $(i+1,j+1)$ and $(i+1,j-1)$, which are the three relevant neighboring sites of the intermediate site $(i+1,j)$ (see Fig. \ref{2d_CBM_fig}). Depending on whether these sites are occupied or empty, the coefficient of the process will be modified accordingly. Essentially there are four cases: 1) all the three NN sites are empty ; 2) any one of the three neighboring sites is occupied ; 3) any two of the NN sites are occupied; and 4) all the three neighboring sites are occupied. Considering all the cases above, we end up with the following 
{NNN and NNNN repulsion terms in $H^{(
2)}$ as detailed in Appendix B.}
\subsubsection{NNN repulsion along diagonals}
The first term is the NNN repulsion which acts along the diagonals of the square lattice; it is given by
\begin{align}
V_2\sum_{i,j}\left(n_{i,j}n_{i+1,j+1}+n_{i,j}n_{i-1,j+1}\right),
\end{align}
where 
\begin{align}
 &V_2=2t^2\Bigg[\left(\frac{1}{2}-m\right)^2\frac{2V_p}{(E_p+V_p)(E_p+2V_p)}\nonumber\\
&\qquad+\left(\frac{1}{4}-m^2\right)\frac{4E_pV_p}{(E_p+V_p)(E_p+2V_p)(E_p+3V_p)}\nonumber\\
&\qquad+\left(\frac{1}{2}+m\right)^2\frac{2E_pV_p}{(E_p+2V_p)(E_p+3V_p)(E_p+4V_p)}\Bigg],
\label{V2}
\end{align}
 with $m$ being the magnetization of the system.
\subsubsection{NNNN repulsion along the x- and y-axes}
We find the second term to be the NNNN repulsion which acts along the x- and y-axes of the square lattice;
it is given by
\begin{align}
V_3\sum_{i,j}\left(n_{i,j}n_{i+2,j}+n_{i,j}n_{i,j+2}\right),
\end{align}
with $V_3=\frac{V_2}{2}$.

{It is important to note that, in the absence of the NN repulsion $2V_p$, we obtain expressions
for $V_z$, $V_2$, and $V_3$ consistent with the non-cooperative treatment of the
electron-phonon interaction in Ref. \onlinecite{kar}.}
\subsection{NNN and NNNN hoppings}\label{subsec_NNN_hopping}
{The remaining terms in $H^{(2)}$ are the hoppings of the HCBs to the NNN and NNNN sites. Similar to
the NNN and NNNN repulsions, the  hopping contributions of the HCBs can also be divided into two types:
NNN hopping along the diagonals and NNNN hopping along the x- and y-axes (see Appendix C for details).}
\subsubsection{NNN hopping along diagonals}
While calculating the coefficient of the NNN hopping, we have to keep in mind the fact that the HCB passes through an intermediate site while hopping to its NNN site. So the coefficient must depend on the occupancy of the two neighboring sites of the intermediate site. For example, if a HCB at site $(i,j)$ is hopping to its right-upper diagonal site, i.e., $(i+1,j+1)$, it can follow any one of the two possible paths: a) first going along x-axis to the $(i+1,j)$-th site and then along y-axis to the $(i+1,j+1)$-th site; and b) the interchanged process, i.e., hopping along the y-axis first to the $(i,j+1)$-th site followed by a hop along the x-axis to the $(i+1,j+1)$-th site (see Fig. \ref{2d_CBM_fig}). For the first path, the coefficient of the hopping depends on whether the two sites $(i+2,j)$ and $(i+1,j-1)$, which are NN of the intermediate site $(i+1,j)$, are occupied or empty. On the other hand, for the second path, the hopping coefficient depends on the occupancy of the two neighboring sites of the 
intermediate 
site $(i,j+1)$, i.e., $(i-1,j+1)$ and $(i,j+2)$. To calculate the NNN hopping coefficient, first we forget about the 
{
occupancy of the two neighbors of the intermediate site;
then, the NNN hopping along the diagonals is obtained to be}
\begin{align}
 -\frac{2t^2e^{-E_p/\omega_0}}{E_p+2V_p}\sum_{i,j}\left(d_{i+1,j+1}^\dagger d_{i,j}+d_{i-1,j+1}^\dagger d_{i,j}+{\rm H.c.}\right),
\end{align}
where 
the coefficient $\frac{2t^2e^{-E_p/\omega_0}}{E_p+2V_p}$ is an approximation with the exact expression
being given in Appendix C.

Now, taking the two neighbors of the intermediate site into account, the NNN hopping term along the diagonals of the square lattice gets  modified to be
\begin{align}
 -t_2\sum_{i,j}\left(d_{i+1,j+1}^\dagger d_{i,j}+d_{i-1,j+1}^\dagger d_{i,j}+{\rm H.c.}\right),
\end{align}
where 
\begin{align}
\!\!\!\! t_2=\frac{2t^2e^{-E_p/\omega_0}}{E_p+2V_p} \Bigg[&\left(\frac{1}{2}-m\right)^2
 +  \left(\frac{1}{4}-m^2\right)\frac{2E_p+4V_p}{E_p+4V_p}  \nonumber\\
 &+\left(\frac{1}{2}+m\right)^2\frac{E_p+2V_p}{E_p+6V_p}\Bigg] .
 \label{t2}
\end{align}
\subsubsection{NNNN hopping along the x- and y-axes}
Next, we consider the hopping of the HCBs to the NNNN sites along the x- and y-axes of the square lattice. 
Similar to the previous case, the coefficient of the hopping in this case, depends on the occupancy of the 
two neighboring sites of the intermediate site. For example, if a HCB is hopping from site $(i,j)$ to its NNNN 
site $(i+2,j)$, it has to pass through the intermediate site $(i+1,j)$ (see Fig. \ref{2d_CBM_fig}). 
So, the coefficient for this process depends on whether the  neighboring sites of site
$(i+1,j)$, i.e., $(i+1,j+1)$ and $(i+1,j-1)$, are occupied or empty. Taking into account 
all the occupancy  possibilities  of the neighboring sites of the intermediate site,
 we get the NNNN hopping term 
to be
\begin{align}
 -t_3\sum_{i,j}\left(d_{i+2,j}^\dagger d_{i,j}+d_{i,j+2}^\dagger d_{i,j}+{\rm H.c.}\right),
\end{align}
with $t_3=\frac{t_2}{2}$.

{Again, it should be pointed out that, in the absence of the NN repulsion $2V_p$, 
the expressions
for $t_2$ and $t_3$ simplify to be consistent with the results of the non-cooperative analysis of the
electron-phonon interaction in Ref. \onlinecite{kar}.}

Finally, taking all the terms present in $H^{(2)}$ into account, $H_{\rm{eff}}$ in Eq. (\ref{H_eff}) reduces to
\begin{align}
H_{\rm{eff}}=&-\left(E_p+2V_z\right)\sum_{i,j}n_{i,j}\nonumber\\
&-t_1\sum_{i,j}\left(d_{i+1,j}^{\dagger}d_{i,j}+d_{i,j+1}^{\dagger}d_{i,j}+{\rm H.c.}\right)\nonumber\\
&+V_1\sum_{i,j}\left(n_{i,j}n_{i+1,j}+n_{i,j}n_{i,j+1}\right)\nonumber\\
&-t_2\sum_{i,j}\left(d_{i+1,j+1}^{\dagger}d_{i,j}+d_{i-1,j+1}^{\dagger}d_{i,j}+{\rm H.c.}\right)\nonumber\\
&+V_2\sum_{i,j}\left(n_{i,j}n_{i+1,j+1}+n_{i,j}n_{i-1,j+1}\right)\nonumber\\
&-t_3\sum_{i,j}\left(d_{i+2,j}^{\dagger}d_{i,j}+d_{i,j+2}^{\dagger}d_{i,j}+{\rm H.c.}\right)\nonumber\\
&+V_3\sum_{i,j}\left(n_{i,j}n_{i+2,j}+n_{i,j}n_{i,j+2}\right),
\label{H_final}
\end{align}
where 
$t_1=te^{-(E_p+V_p)/\omega_0}$, $~V_1=2V_p+V_z$,
and 
{the expressions for all the remaining terms, $V_z$, $t_2$, $t_3$, $V_2$, and $V_3$, 
being the same as defined earlier.}

\section{Numerical Calculations}\label{Numerical_calc}
To study the phase diagram of our effective Hamiltonian of HCBs, we use quantum Monte Carlo (QMC) simulation employing the stochastic-series-expansion (SSE) technique. The first step required for SSE is to rewrite the Hamiltonian in terms of spin-1/2 operators. Identifying the relations between the operators for HCBs and those for spin-1/2 particles as 
 $d_{i,j}^\dag=S_{i,j}^+$, $d_{i,j}=S_{i,j}^-$ and $n_{i,j}=S_{i,j}^z+\frac{1}{2}$,
we recast our effective Hamiltonian for HCBs,
in units of $2t_1$, as an extended XXZ spin-1/2 Hamiltonian, given by
\begin{align}
H&=\sum_{i,j}\Big[-\frac{1}{2}\left(S_{i+1,j}^+S_{i,j}^-+S_{i,j+1}^+S_{i,j}^-+{\rm H.c.}\right)\nonumber\\
&\quad\quad\quad\;+\Delta_{1}\left( S_{i,j}^z S_{i+1,j}^z+S_{i,j}^z S_{i,j+1}^z\right)\Big]\nonumber\\
&+\sum_{i,j}\Big[-\frac{J_{2}}{2}\left(S_{i+1,j+1}^+S_{i,j}^-+S_{i-1,j+1}^+S_{i,j}^-+{\rm H.c.}\right)\nonumber\\
&\quad\quad\quad\;+\Delta_{2}\left( S_{i,j}^z S_{i+1,j+1}^z+S_{i,j}^z S_{i-1,j+1}^z\right)\Big]\nonumber\\
&+\sum_{i,j}\Big[-\frac{J_{3}}{2}\left(S_{i+2,j}^+S_{i,j}^-+S_{i,j+2}^+S_{i,j}^-+{\rm H.c.}\right)\nonumber\\
&\quad\quad\quad\;+\Delta_{3}\left( S_{i,j}^z S_{i+2,j}^z+S_{i,j}^z S_{i,j+2}^z\right)\Big]\nonumber\\
&-h_0\sum_{i,j} S_{i,j}^z.
\label{H_SSE}
\end{align}
Looking at Eqs. (\ref{H_final}) and (\ref{H_SSE}), one can easily see that 
$J_{2}=t_2/t_1$, $J_{3}=t_3/t_1$, $\Delta_{1}=V_1/(2t_1)$,
$\Delta_{2}=V_2/(2t_1)$, $\Delta_{3}=V_3/(2t_1)$ and $h_0=E_p+2V_z-2V_1-2V_2-2V_3$;
here, $J_i$ and $\Delta_i$ are the transverse and longitudinal couplings, respectively.

Now, to figure out the phase diagram of the system, we need to study the Hamiltonian at various filling-fractions of HCBs. To vary the number of HCBs in the system, or in other words to tune the magnetization of the spin-1/2 system, we  replace the constant $h_0$ by a variable $h$ in the term $-h_0\sum\limits_{i,j} S_{i,j}^z$
of the Hamiltonian $H$ given by Eq. (\ref{H_SSE}); 
here $h$ is taken as the external magnetic field in units of $2t_1$. By tuning the external magnetic field $h$, we can actually tune the magnetization of the system and study the behavior of the system at various fillings.

We use two kinds of order parameter: structure factor $S(\vec{Q})$ (to identify diagonal long-range
order) and superfluid density $\rho_s$ (to identify off-diagonal long-range order)
and construct the phase diagram. The structure factor per site is defined as 
\begin{align}
S(\vec{Q})=\frac{1}{N_s^2}\sum_{i,j}\sum_{m,n}e^{i\vec{Q}\cdot(\vec{R}_{i,j}-\vec{R}_{m,n})}\langle S_{i,j}^z S_{m,n}^z\rangle,
\end{align}
with $\langle...\rangle$ being the ensemble average. We study $S(\vec{Q})$ at all values of $\vec{Q}$ and identify those that produce peaks in the structure factor. Here we would like to point out that the maximum possible value of $S(\vec{Q})$ is $0.25$.

The superfluid density is expressed in terms of the winding numbers, $W_x$ and $W_y$, in the x- and y-directions as\cite{sandvik_review} 
\begin{align}
 \rho_s=\frac{1}{2\beta}\langle W_x^2+W_y^2\rangle.
\end{align}
The winding number $W_x$ along the x-direction can be calculated as $W_x=\frac{1}{L_x}(N_x^+-N_x^-)$, where $N_x^+$ and $N_x^-$ denote the total number of operators transporting spin in positive and negative x-directions, respectively and $L_x$ denotes the length of the lattice along the x-direction.
\begin{table}[t]
\begin{tabular}{|c|c|c|c|}
\hline
 $\tilde{g}$ & $\Delta_1$ & $(\Delta_2)_{max}$ & $(J_2)_{max}$\\ \hline
 $1.0$ & $1.7436$ & $0.4757$ & $1.6486$\\ \hline
 $1.5$ & $5.7744$ & $0.7379$ & $0.8760$\\ \hline
 $1.8$ & $16.6463$ & $1.3791$ & $0.7007$\\ \hline
 $2.0$ & $39.2161$ & $2.3887$ & $0.6327$\\ \hline
 $2.25$ & $131.8584$ & $5.4612$ & $0.5818$\\ \hline
 $2.5$ & $507.9968$ & $14.5044$ & $0.5584$\\ \hline
 $3.0$ & $10896.8217$ & $157.5599$ & $0.5744$ \\ \hline
\end{tabular}
\caption{Values of NN longitudinal coupling $\Delta_1$ and maximum values of NNN longitudinal coupling
$\Delta_2$ and NNN transverse coupling $J_2$ for different values of $\tilde{g}$. }\label{table1}
\end{table}

\begin{figure}[b]
\includegraphics[width=0.9\linewidth,angle=0]{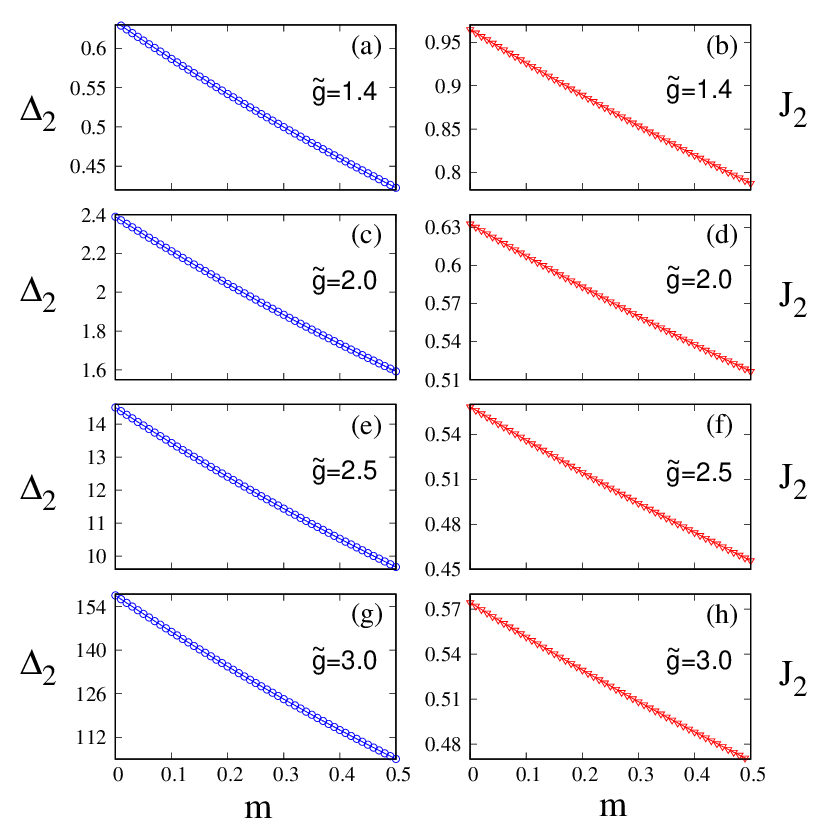}
\caption{
(Color online) Dependence of NNN longitudinal coupling 
$\Delta_2$ and NNN transverse coupling
$J_2$ on magnetization $m$ as derived from Eqs. (\ref{V2}), (\ref{t2}), (\ref{H_final}), and (\ref{H_SSE})
for the following cases:
(a) \& (b) at $\tilde{g}=1.4$; (c) \& (d) at $\tilde{g}=2.0$; (e) \& (f) at $\tilde{g}=2.5$; and (g) \& (h) at $\tilde{g}=3.0$.
}
\label{Delta2J2}
\end{figure}

\begin{figure}[b]
    \centering
    \begin{minipage}[b]{0.18\textwidth}
        \includegraphics[width=\textwidth]{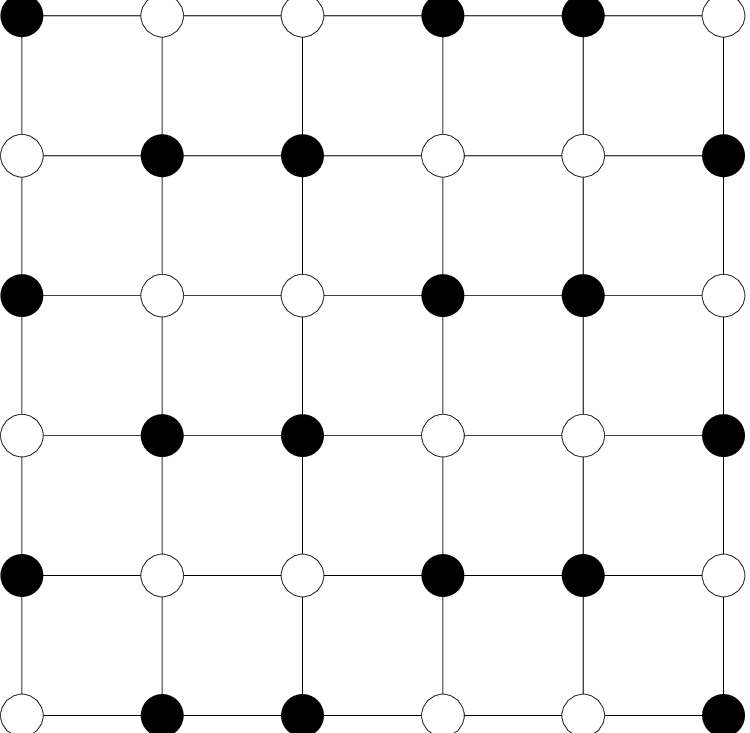}
        \vskip 1ex (a)
    \end{minipage}
 \quad\quad
    \begin{minipage}[b]{0.18\textwidth}
        \includegraphics[width=\textwidth]{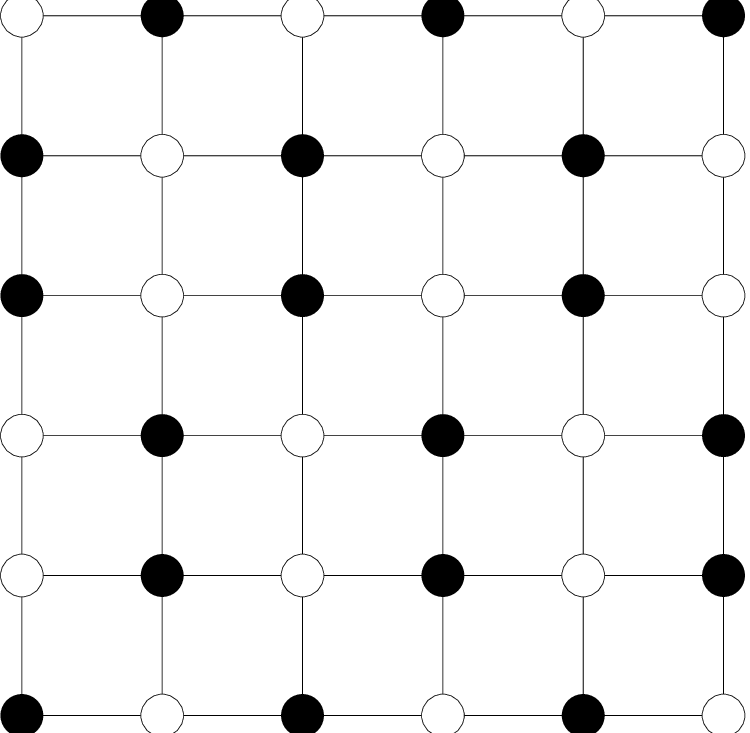}
         \vskip 1ex (b)
    \end{minipage}
    \caption{
Two types of honeycomb-like solid depicted by a peak in (a) $S(\pi/2,\pi)$ and (b) $S(\pi,\pi/2)$}\label{fig:honeycomb}
\end{figure}

We now discuss the values of different parameters in our Hamiltonian given by Eq. (\ref{H_SSE})
and used in our numerical calculations. We concentrate on the case $t/\omega_0=1.0$ for the construction of our phase diagram.
Since $\gamma=\sqrt{2}$, we set $\tilde{g}^2=7g^2$ so as to get the simple expression 
$E_p+V_p = \tilde{g}^2 \omega_0$.
The coefficients $J_2~ (=2J_3)$ and $\Delta_2~(=2\Delta_3)$ depend on the magnetization $m$ of the system. While Fig. \ref{Delta2J2}
depicts that $J_2$ and $\Delta_2$ values (at various couplings $\tilde{g}$)
monotonically decreases with increasing magnetization $m$,
Table \ref{table1} shows the values of $\Delta_1$ and the maximum values of $\Delta_2$ and $J_2$ 
for different values of $\tilde{g}$. {
As one can see,  $\Delta_1 / (\Delta_2)_{\rm max}$ increases monotonically
approximately from $3.665$ to $69.159$ as $\tilde{g}$ is varied from $1.0$ to $3.0$. 
At larger values of $\tilde{g}$, when $\Delta_1$ and $\Delta_2$ assume large values,
our numerical calculations suffer from significant slowing down resembling the
situation in Ref. \onlinecite{kar}; with our computational constraints
we cannot use exact values when $\Delta_1$ and $\Delta_2$ assume large values.
We can set a cut-off for the  parameters $\Delta_1$ and $\Delta_2$
above which the essential physics for our system remains unaltered.
Similar to Ref. \onlinecite{kar}, the upper cut-off for $\Delta_1$ is 16.
Furthermore, to identify the cut-off for $\Delta_2$,
we need to find out the lowest value of $ \Delta_1/\Delta_2 $
which can be used without changing the essential physics. To this end,
we have calculated the superfluid density and structure factor at half-filling 
\big (where $\Delta_2 = (\Delta_2)_{\rm max} $\big ) for the following 
set of values of $\big (\Delta_1,(\Delta_2)_{\rm max}\big )$: $(20,4)$, $(20,5)$, $(20,6)$, $(16,5)$,
$(20,7)$, $(17,6)$, $(16,7)$, and $(20,9)$ with the value of $\Delta_1 / (\Delta_2)_{\rm max}$ being
$5$, $4$, $3.33$, $3.2$, $2.86$, $2.83$, $2.29$,  and $2.22$, respectively.
Numerical results show that for the first four cases, 
where $\Delta_1 > 3 (\Delta_2)_{\rm max}$, at half-filling the system manifests a checkerboard 
solid (cS)
with a peak in the structure factor $S(\pi,\pi)$. On the other hand, for the last four cases where $2(\Delta_2)_{\rm max}<\Delta_1<3(\Delta_2)_{\rm max}$, at half-filling the system produces
a completely different type of solid depicted in Fig. \ref{fig:honeycomb} (which we call honeycomb-like solid), 
indicated by a peak in $S(\pi/2,\pi)$ or $S(\pi,\pi/2)$. The reason can be explained as follows. 
In the cS phase each particle feels $6(\Delta_2)_{\rm max}$ amount repulsion, whereas in the honeycomb-like solid the repulsion felt by each particle is $\Delta_1+3(\Delta_2)_{\rm max}$. 
The checkerboard solid will be favored over the honeycomb-like solid only if
$\Delta_1+3(\Delta_2)_{\rm max}>6(\Delta_2)_{\rm max}$,
i.e., $\Delta_1>3(\Delta_2)_{\rm max}$. Therefore to capture the correct physics of our 
system, the minimum value of $ \Delta_1/\Delta_2 $
must be greater than $3$. Keeping all these facts in mind, we set the cut-off values to be $\Delta_1=16$ and $\Delta_2=5$
(with $\Delta_3=\frac{\Delta_2}{2}$), so that the physics of the system still remains the same.}

All numerical results in Figs. \ref{g_1.4_2.5}--\ref{min_ss}
have been 
obtained {
in a $18\times18$ lattice with $t/ \omega_0=1.0$}.

\section{Results and Discussions}\label{results}
To determine the various phases of our 2D $t_1-t_2-t_3-V_1-V_2-V_3$ model,
one needs to understand the interplay between different types of hopping and repulsion.
To construct the phase diagram, we vary the magnetization $m$ from $0$ to $0.5$;
this corresponds to varying the 
particle filling $\rho$ from $1/2$ to 1.
Due to particle-hole symmetry of the Hamiltonian, the physics at any filling-fraction 
for  particles  is identical to that for holes at the same filling.

\begin{figure}[b]
\includegraphics[width=0.9\linewidth,angle=0]{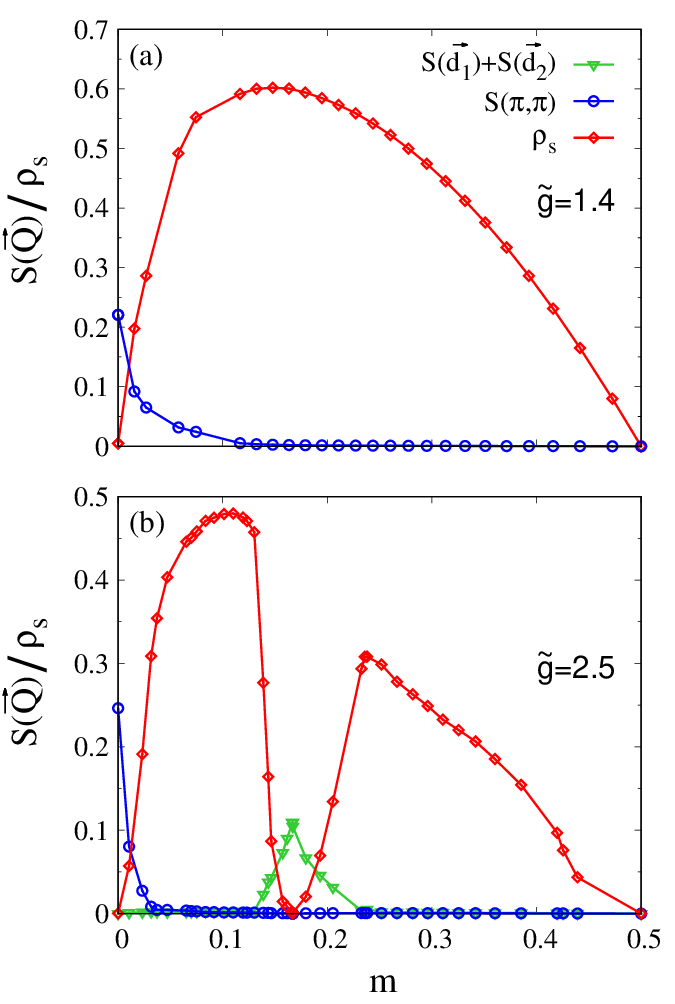}
\caption{
(Color online) Plots of structure factor $S(\vec{Q})$ and superfluid density
$\rho_s$ vs magnetization $m$ 
for HCBs on  a $18\times18$ lattice  with $t/\omega_0=1.0$ and when
(a) $\tilde{g}=1.4$ and (b) $\tilde{g}=2.5$.
{Curves are averaged results from simulations using three different random number seeds.}
}
\label{g_1.4_2.5}
\end{figure}

\begin{figure}[t]
 \centering
 \begin{minipage}[c]{0.46\textwidth }
  \includegraphics[width=0.4\columnwidth]{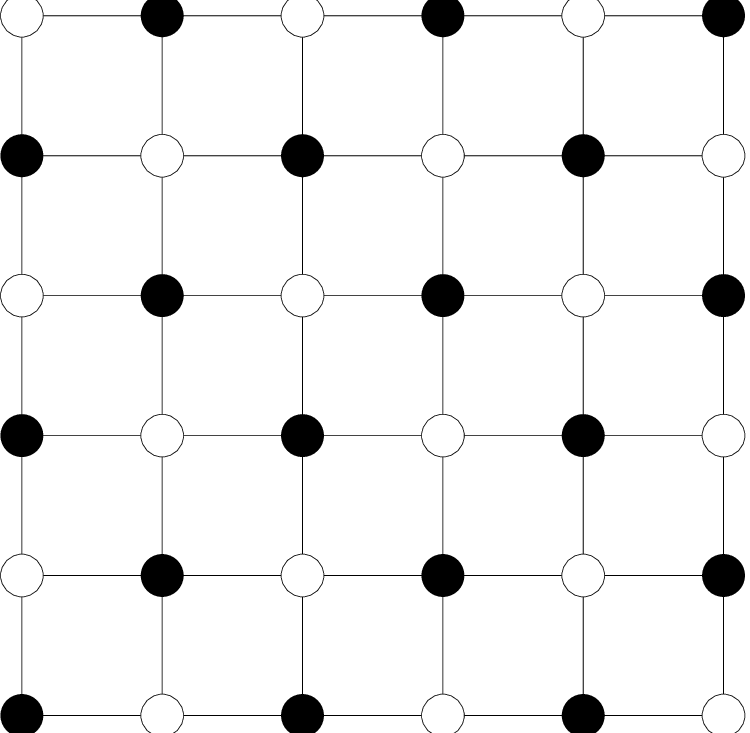}
  \vskip 1ex (a)
  \end{minipage}
  \vskip 2ex
  \begin{minipage}[c]{0.18\textwidth}
        \includegraphics[width=\textwidth]{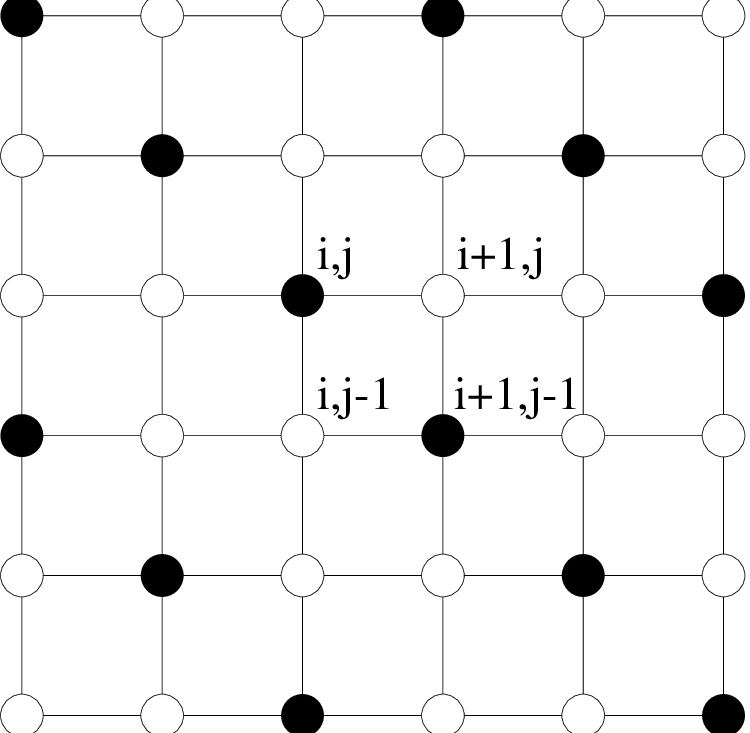}
            \vskip 1ex  (b)
            \end{minipage}
            \quad\quad
      \begin{minipage}[c]{0.18\textwidth}
        \includegraphics[width=\textwidth]{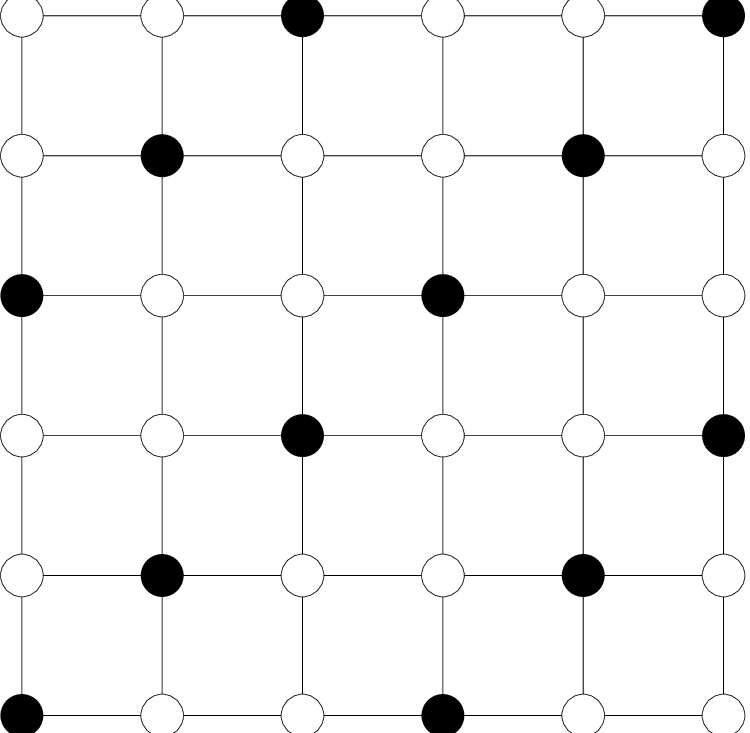}
               \vskip 1ex (c)
      \end{minipage}            
 \caption{Different types of CDWs: (a) checkerboard solid (cS) at half-filling with $S(\vec{Q})$ peaking at 
    $\vec{Q} =(\pi,\pi)$; (b) diagonal striped solid (dsS) indicated by peak in $S(\vec{Q})$ at 
    $\vec{Q}=(2\pi/3,2\pi/3)$; and (c) dsS characterized by ordering wavevector $\vec{Q} = (2\pi/3,4\pi/3)$.
       } \label{fig:lattice}   
\end{figure}

Figure \ref{g_1.4_2.5} shows the variation of the structure factor $S(\vec{Q})$ and 
the superfluid density $\rho_s$ as a function of the magnetization $m$, for two different values of $\tilde{g}$, i.e., $1.4$ and $2.5$. A key point to note here is that, in general,
larger values of repulsion aid in the formation of a CDW, whereas larger values of 
NNN tunneling $t_2$ help a particle hop  in the same sublattice.
For $\tilde{g}=1.4$, at half-filling, the HCBs
form a checkerboard solid shown in Fig. \ref{fig:lattice}(a) and  indicated by a peak in the structure 
factor $S(\pi,\pi)$. Slightly away from half-filling,  a supersolid region develops after which the system 
retains only its superfluidity. The reason can be understood by examining the 
coefficients of different terms in the Hamiltonian in Eq. (\ref{H_SSE}). Since the NN repulsion dominates over the NNN 
and NNNN repulsions, at half-filling the system becomes a cS phase to avoid NN occupation, even though the particles experience NNN and NNNN repulsions. Now, if we add one additional particle to the half-filled system, 
the extra particle can be at any one of the empty sites; irrespective of the site it resides on, the particle will
feel the same  extra repulsion  $4V_1$. This extra particle can hop to its NNN or NNNN sites, 
without changing the repulsive interaction in the system which has a checkerboard solid in the background,
resulting in the coexistence of superfluidity 
and 
CDW state. If we keep 
on increasing the particle number, after a certain filling-fraction, the checkerboard structure is lost with the system continuing to be a superfluid.

\begin{figure}[b]
\includegraphics[width=1.\linewidth,angle=0]{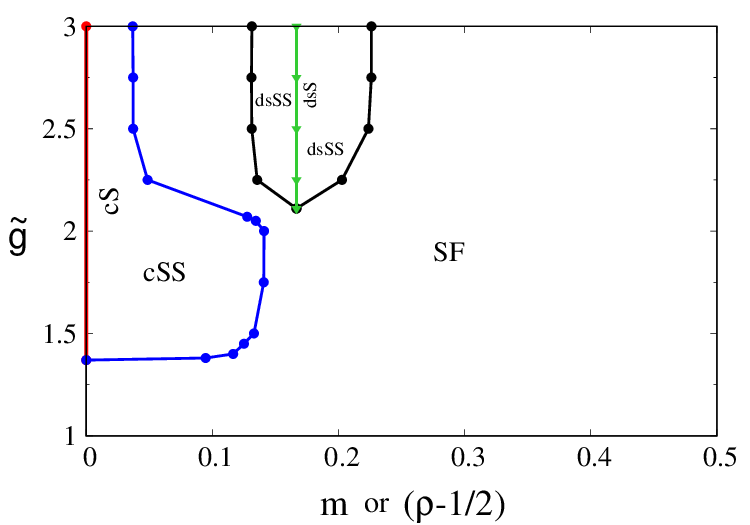}
\caption{
(Color online) Phase diagram in terms of magnetization (or filling-fraction $\rho$)
for HCBs on  a $18\times18$ lattice  with $t/\omega_0=1.0$.
cS represents checkerboard solid with cSS being the corresponding supersolid;
dsS stands for diagonal striped solid with dsSS being the related supersolid.
{Plots represent averaged results from simulations employing three different random number seeds}.
}
\label{phase_diagram}
\end{figure}

\begin{figure}[t]
\includegraphics[width=0.9\linewidth,angle=0]{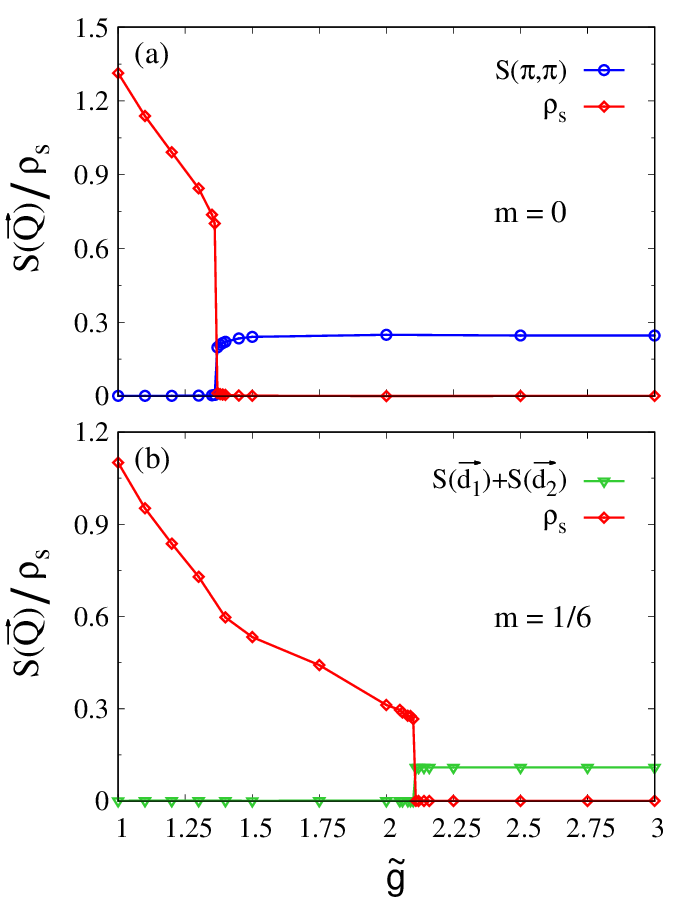}
\caption{
(Color online)  Plots of $S(\vec{Q})$ and $\rho_s$ vs coupling strength $\tilde{g}$ 
depicting first-order transitions at two different magnetization values: (a) $m=0$ (or  half-filling) and (b) $m=1/6$ (or two-third filling).}
\label{1st_order}
\end{figure}

\begin{figure}[b]
\includegraphics[width=1\linewidth,angle=0]{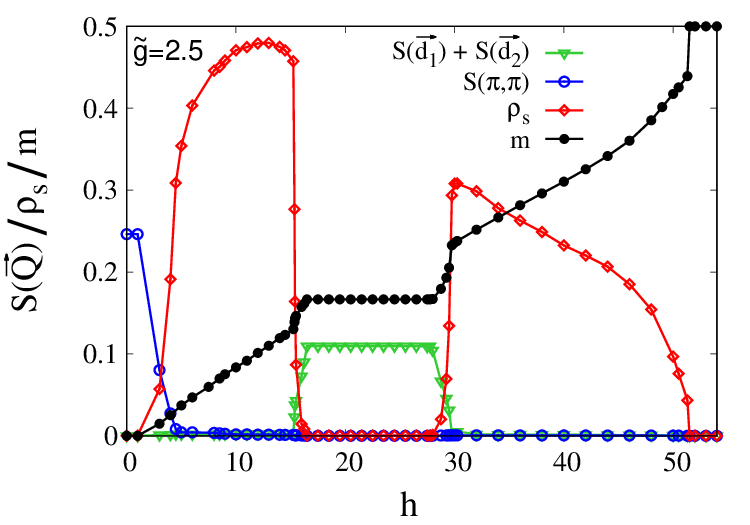}
\caption{
(Color online) Evolution of order parameters $S(\vec{Q})$, $\rho_s$ and $m$ as the magnetic field $h$ is varied at a fixed coupling strength $\tilde{g}=2.5$. 
{No discontinuous transitions are exhibited}.
}
\label{fig_field}
\end{figure}

\begin{figure}[t]
\includegraphics[width=0.9\linewidth,angle=0]{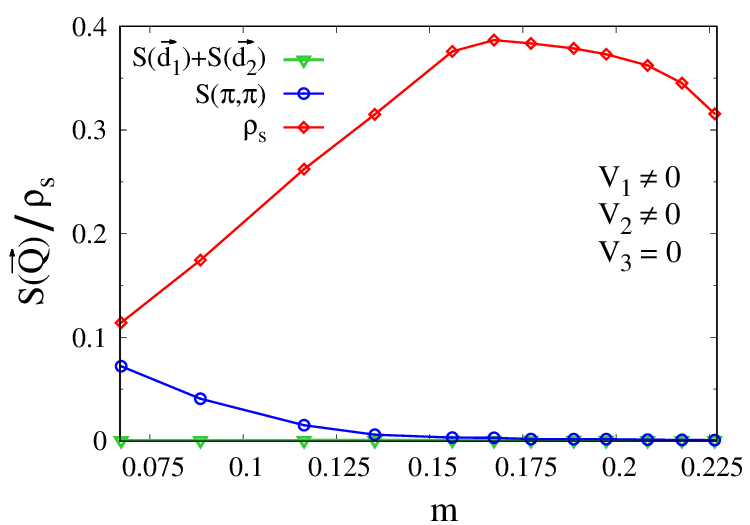}
\caption{
(Color online) Variation of $S(\vec{Q})$ and $\rho_s$ vs magnetization $m$ in the absence of 
the NNNN repulsion $V_3$ along x- and y-axes
 in the $t_1-t_2-t_3-V_1-V_2-V_3$ model of Eq. (\ref{H_final}).\\
}
\label{fig_delta}
\end{figure}

\begin{figure}[b]
\includegraphics[width=1\linewidth,angle=0]{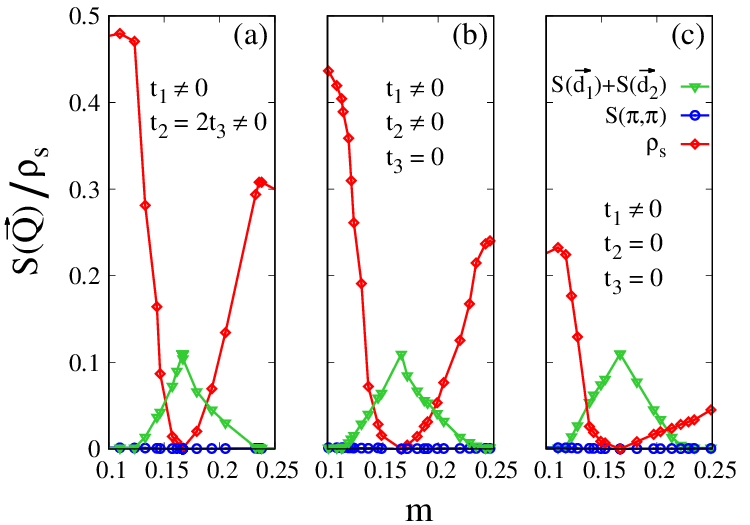}
\caption{
(Color online) Plots of $S(\vec{Q})$ and $\rho_s$ vs magnetization $m$, in the vicinity of striped phase, for three different cases
in the $t_1-t_2-t_3-V_1-V_2-V_3$ model of Eq. (\ref{H_final}): (a) all the three hoppings $t_1$, $t_2$, and $t_3$
are present; (b) NNNN hopping $t_3$
along x- and y-axes is set to zero; and (c) only NN hopping $t_1$ is present.
The minimum model for diagonal striped solid (dsS) is shown to be $t_1-V_1-V_2-V_3$.
}
\label{min_ss}
\end{figure}

Now looking at Fig. \ref{g_1.4_2.5}(b) for $\tilde{g}=2.5$, we see that an additional CDW appears at 
fillings $\rho =1/3~ {\rm and}~ 2/3$. Since
the physics pertinent to $\rho =1/3$ is the same as that for $\rho =  2/3$, we will analyze
 them interchangeably based on our convenience.
At $\rho = 1/3$,
the HCBs form a diagonal striped solid manifesting spontaneously broken  symmetry and
characterized by a peak in the structure factor at wavevector {
 $\vec{d_1}=(2\pi/3,2\pi/3)$ 
[corresponding to Fig. \ref{fig:lattice}(b)] or $\vec{d_2}=(2\pi/3,4\pi/3)$}
[related to Fig. \ref{fig:lattice}(c)]. Although each particle in the stripe experiences a repulsion $2V_2$,
it is still the minimum energy state of the system at one-third filling. If we add one extra particle to the system,
it occupies any one of the empty sites between the stripes and experiences a repulsion $2V_1+V_2+2V_3$. 
Now, this extra particle can hop to any of its unoccupied NN, NNN, or NNNN sites without a change in the potential energy of the system;
thus, coexistence of stripe order and superfluidity is realized on the interstitial side. 
On the other hand, if we remove
one particle from the system at $\rho=1/3$, the extra hole (residing in the stripes) can hop 
along the stripes without  altering the potential energy; thus, supersolidity is exhibited
on the other (i.e., vacancy) side of the diagonal striped phase as well.
Thus, the mechanism governing the existence of a supersolid phase 
away from commensurate fillings $1/2$ and $1/3$,  on our unfrustrated system (i.e.,  the square lattice), 
is that interstitials or vacancies can move without frustration, i.e., without
a cost in the potential energy.

The complete ground state phase diagram is depicted in Fig. \ref{phase_diagram}.
The half-filled system shows the signature of a checkerboard solid (cS) for all $\tilde{g}$ values
above $1.37$. Next to this CDW, we have a supersolid region (cSS) 
where $S(\pi,\pi)$ and $\rho_s$ coexist homogeneously.
On the other hand, at filling fraction $\rho=1/3$, the system realizes a dsS
beyond $\tilde{g} = 2.11$. On both sides of this striped solid, 
we have a region of supersolid (dsSS) which is a homogeneous coexistence of the diagonal striped
solid and a superfluid. 
As we increase $\tilde{g}$ beyond $1.37$,
the width of the supersolid region cSS  increases and attains its maximum 
at $\tilde{g}=2.0$. Further increase in $\tilde{g}$ results in a decrease in the width of the cSS region,
thereby making way for diagonal stripe supersolid  at higher values of $\tilde{g}$.
However, we should point out that there is no direct supersolid-supersolid transition.
Next, it is interesting to note that there is an asymmetry
in the extent of the dsSS region around one-third filling. Thus,
there is an asymmetry at $\rho=1/3$
with respect to doping with interstitials and vacancies similar
to the asymmetry at one-fourth filling reported in Ref. \onlinecite{boninsegni1} for a
$t_1-V_1-V_2$ model when $V_1 < 2V_2$.
It is also worth noting that, at lower fillings such as $\rho=1/4$ and $\rho=1/5$,
there is no CDW order.

In a recent study of HCBs on a square lattice with NN hopping and NN repulsion (i.e., in a $t_1-V_1$ model),
when a sizeable external potential is applied along the diagonal stripes 
in Figs. \ref{fig:lattice}(b) and \ref{fig:lattice}(c), 
the authors obtain the corresponding diagonal striped CDW at $\rho=1/3$ and a striped supersolid phase away from
 one-third filling \cite{stripe_long_dang}.
Similar to our case, the physics governing the formation of a supersolid phase
is  that the interstitial particles or vacancies 
in the vicinity of the  commensurate filling $\rho = 1/3$ can hop without changing the potential energy of the system.

{
 In our simulations using SSE, we cannot tune the magnetization (density) directly. 
Instead, we tune the magnetic field which determines the magnetization of the system. For a particular 
value of the magnetic field, the resulting magnetization generally fluctuates during simulation. 
As a result, usually it is not possible to study the nature of the phase transitions
by keeping the magnetization (filling-fraction) fixed at a particular value
and varying $\tilde{g}$. However, when the system is in a CDW state, the magnetization remains 
constant over a range of magnetic field values; this makes it possible to vary  $\tilde{g}$ 
at a fixed magnetization.

We see from Fig. \ref{1st_order}(a) that for the half-filled system (i.e., at $m=0)$, as we increase 
the $\tilde{g}$ value from 1 to 3, the structure factor $S(\pi,\pi)$ jumps from $0$ to
almost its maximum value  and the superfluid density suddenly
drops down to zero at $\tilde{g}=1.37$.
In the phase diagram (depicted in Fig \ref{phase_diagram}), 
this indicates  a first-order transition at $\tilde{g}=1.37$
from a  superfluid to a  checkerboard solid at filling-fraction $\frac{1}{2}$; since
the transition is from a U(1) symmetry breaking state to a  translational symmetry breaking state,
the order of the transition is consistent with Landau's picture.
Similarly at magnetization $m=\frac{1}{6}$ corresponding to filling fraction $\frac{2}{3}$, 
at $\tilde{g}=2.11$, Fig. \ref{1st_order}(b) shows a dramatic jump in the structure factor $S(2\pi/3,2\pi/3)+S(2\pi/3,4\pi/3)$ from 0 to its maximum value accompanied by a discontinuous
drop in the superfluid density to zero. This signifies a first-order  transition
as we move along the $\tilde{g}$-axis at $m = 1/6$ in the phase diagram (shown in Fig \ref{phase_diagram}).
Thus, consistent with the literature \cite{boninsegni1,wessel1,wessel4}, no supersolidity is detected at commensurate fillings in our
unfrustrated system.

In contrast to Fig. \ref{1st_order}(a), Fig. (\ref{fig_field}) depicts the behavior of the order parameters (i.e., structure factor, superfluid density and magnetization)  as we tune the magnetic field
at the fixed value of coupling $\tilde{g}=2.5$. The continuous change in
the order parameters as a function of magnetic field $h$ eliminates
the possibility of a first-order phase transition. This further signifies that in the phase diagram (displayed in Fig \ref{phase_diagram}), as we move along the $m$-axis at any particular $\tilde{g}$ value, all the different phases are separated from each other via continuous phase transitions, i.e.,
all supersolid-solid and superfluid-supersolid transitions are second order.

We will now identify the minimum model for the diagonal striped supersolid.
Compared to the checkerboard supersolid,
the dsSS phase is rarely observed. 
To determine the minimum model for the realization of the dsSS phase, 
we first  identify the necessary repulsions required to observe the diagonal striped solid phase
in the $t_1-t_2-t_3-V_1-V_2-V_3$ model of Eq. (\ref{H_final}). 
From Fig. (\ref{fig_delta}), we see that, as soon as we tune the NNNN repulsion $V_3$
along x and y-axes to zero, the structure factor corresponding to the dsS phase completely disappears. This feature can be explained based on 
Figs. \ref{fig:lattice}(b) and \ref{fig:lattice}(c). For instance, when the NNNN repulsion $V_3$
is set to zero in the structure given by Fig. \ref{fig:lattice}(b), the particles at sites (i,j) and (i+1,j-1)
can both be shifted to
 the neighboring sites (i+1,j) and (i,j-1)  without changing the potential energy of the system.
This process destroys the striped structure. Thus, it follows that all the three repulsions
(i.e., $V_1$, $V_2$, and $V_3$) 
are necessary to stabilize the dsS structure. A similar argument
can be made to destroy the structure given by Fig. \ref{fig:lattice}(c).

Next, in Fig. \ref{min_ss}, we focus on the region in the vicinity of the striped phase.
Compared to Fig. \ref{min_ss}(a), in which all the three hopping parameters are non-zero,
the superfluid density reduces slightly when the NNNN hopping $t_3$ is set to zero [as can be seen in
Fig. \ref{min_ss}(b)]. The interesting feature to note is that, even when only NN hopping $t_1$ is present with the other two hopping parameters $t_2$ and $t_3$ being zero [as in Fig. \ref{min_ss}(c)], we have a 
diagonal striped supersolid region around $m=1/6$ with the width of the dsSS being almost unaffected. 
This elucidates the fact that the minimum model to obtain a dsSS phase is the $t_1-V_1-V_2-V_3$ model.
}

\section{Comparison with LSNO experimental results}\label{comparison}
Stripe-like charge order has been reported in a number of layered transition-metal oxides \cite{stripe_review}.
Among these compounds, the layered nickelate LSNO is an archetypal system to exhibit a
firm charge stripe order. In $\rm{La_{2-x}Sr_x NiO_4}$, static checkerboard charge
order [such as in Fig. \ref{fig:lattice}(a)]
is expressed at $x=1/2$ and static
diagonal stripe order [as shown in Figs. \ref{fig:lattice}(b) and \ref{fig:lattice}(c)] is manifested at
$x=1/3$  with the transition temperatures at these dopings showing  local maxima
\cite{cheong0, cheong, tokura2, tokura3, tokura4,freeman}.
The observed lattice constant  ratio c/a in LSNO displays a maximum  at
$x=1/2$, thereby indicating that in the region $0 < x < 1/2$
holes are  doped into the $d_{x^2-y^2}$ orbitals and  in the region $1/2 < x < 1$
  holes are doped into the $d_{z^2}$  orbitals \cite{cava, tokura2}.
Measurements of Hall coefficient for $\rm{La_{2-x}Sr_x NiO_4}$
 by T. Katsufuji {\it et al.} \cite{Hall_seebeck}, revealed that the charge carriers change 
from electron-like to hole-like while going from the hole density $x<1/3$ to $x>1/3$.

In the undoped ${\rm La_2NiO_4}$, the oxidation state of nickel is ${\rm Ni^{2+}}$ with the electronic configuration ${\rm [Ar]4s^03d^8}$. Hence, only  $d_{z^2}$ and $d_{x^2-y^2}$ orbitals are relevant in the
doped compound $\rm{La_{2-x}Sr_xNiO_4}$.
The electron-phonon interaction term of the Hamiltonian is given by 
\begin{widetext}
\begin{eqnarray}
\!\!\!
H_{ep} =
-\frac
{ g \omega_0}{4}
\sum_{i,j}
\begin{pmatrix}
d^\dagger_{z^2;i,j}~d^\dagger_{x^2-y^2;i,j}
\end{pmatrix}
\!
\begin{pmatrix}
q_{x;i,j}+q_{y;i,j}+4q_{z;i,j} & -\sqrt{3}q_{x;i,j}+\sqrt{3}q_{y;i,j} \\
-\sqrt{3}q_{x;i,j}+\sqrt{3}q_{y;i,j} & 3q_{x;i,j}+3q_{y;i,j}
\end{pmatrix}
\!
\begin{pmatrix}
d_{z^2;i,j} \\
d_{x^2-y^2;i,j}
\end{pmatrix} ,
\nonumber \\
\label{eq:gen_elph_com}
\end{eqnarray}
\end{widetext}
where the distortions $q_{x;i,j} \equiv (a_{x;i,j}^\dagger+a_{x;i,j})-(a_{x;i-1,j}^\dagger+a_{x;i-1,j})$,
$q_{y;i,j} \equiv (b_{y;i,j}^\dagger+b_{y;i,j})-(b_{y;i,j-1}^\dagger+b_{y;i,j-1})$, and
$q_{z;i,j} \equiv (c_{z;i,j}^\dagger+c_{z;i,j})$.
In the undoped compound, since both  $d_{z^2}$ and $d_{x^2-y^2}$ orbitals are occupied,
there are only breathing mode distortions ($4q_{x;i,j}+4q_{y;i,j}+4q_{z;i,j}$) and
no active Jahn-Teller (JT) distortions. Now, when we introduce holes in the system (by doping with ${\rm Sr}$ such that $0< x< 1/2$), the holes occupy the $d_{x^2-y^2}$ orbitals; this is
because a site with a single electron in $d_{x^2-y^2}$ orbital will
produce in-plane distortions ($3q_{x}+3q_{y}$) which have a
greater incompatibility with the breathing mode distortions ($4q_{x}+4q_{y}+4q_{z}$)
on the adjacent sites and thus
cost more energy than a singly occupied $d_{z^2}$ orbital. 
These $d_{x^2-y^2}$ holes can hop  and are responsible for the transport properties.
Each site with a $d_{x^2-y^2}$ hole is JT active.

 The Hamiltonian for cooperative Jahn-Teller (CJT) distortions in 
 the two-dimensional LSNO system involves holes in $d_{x^2-y^2}$ orbitals as the active  carriers. 
 The starting Hamiltonian $H_{\rm LSNO}$, describing $\rm{La_{2-x}Sr_xNiO_4}$ for $0\leq {\rm x}\leq 0.5$,
 consists of the following terms expressed in terms of the creation (destruction) operator $h^\dagger_{i,j}(h_{i,j})$ for the holes in  $d_{x^2-y^2}$ orbitals.\\
(i) Hopping term 
\begin{align}
 H^\prime_t=\frac{3t}{4}\sum_{i,j}&\big(h_{i+1,j}^\dagger h_{i,j}+h_{i,j+1}^\dagger h_{i,j}+{\rm H.c.}\big);
\end{align}
(ii) hole-phonon interaction term:
\begin{align}
 H^\prime_{I}=\frac{3}{4}g\omega_0&\sum_{i,j}\Big[(a_{x;i,j}^\dagger+a_{x;i,j})(n_{i,j}^h-n_{i+1,j}^h)\nonumber\\
&+(b_{y;i,j}^\dagger+b_{y;i,j})(n_{i,j}^h-n_{i,j+1}^h)\Big] ;
\end{align}
and
(iii) lattice term:
\begin{align}
 H^\prime_l=\omega_0\sum_{i,j}\left(a_{x;i,j}^\dagger a_{x;i,j}+b_{y;i,j}^\dagger b_{y;i,j}\right),
\end{align}
where $n_{i,j}^h \equiv h^\dagger_{i,j} h_{i,j}$.

The Lang-Firsov transformed Hamiltonian is given by $\tilde{H}_{\rm LSNO}={\rm exp}(S)~H_{\rm LSNO}~{\rm exp}(-S)$ where $S$ has the form
\begin{align}
 S=\frac{3}{4}g\sum_{i,j}&\Big[(a_{x;i,j}^\dagger-a_{x;i,j})(n_{i,j}^h-n_{i+1,j}^h)\nonumber\\
 &+(b_{y;i,j}^\dagger-b_{y;i,j})(n_{i,j}^h-n_{i,j+1}^h)\Big] .
\end{align}

Setting $t^\prime=-3t/4$ and $g^\prime=-3g/4$, in the non-adiabatic regime ($|t^\prime|/\omega_0 \leqslant 1$) and at strong coupling (i.e., large ${g^\prime}^2$), the transformed Hamiltonian can be split into 
two terms: the unperturbed Hamiltonian and the perturbation term. These two terms are the same as the ones
given by Eqs. (\ref{unperturbed}) and (\ref{perturbed}), except that they are written in hole-operator language; 
 both $\gamma$ and $\eta$ are set to zero value; and $t$ and $g$ are replaced by 
 $t^\prime$ and $g^\prime$, respectively. 
If the carriers are taken to be HCBs instead of fermionic holes,
then after following  the same second-order perturbative procedure as in Sec. \ref{Eff_Hamiltonian}, 
we end up with an effective Hamiltonian that is exactly the same as that given by Eq. (\ref{H_final})
with $\gamma=0=\eta$ and with
$t$ and $g$ being replaced by $t^\prime$ and $g^\prime$, respectively.
It is important to note that the small parameter value is again given by $\left[\frac{t^2}{2(E_p+V_p)\omega_0}\right]^{\frac{1}{2}}$ and remains unaltered. Now, since we are dealing with
fermionic holes and not HCBs, we get the following effective Hamiltonian:
\begin{align}
H_{\rm{eff}}^{\prime}=&-\left(E_p+2V_z\right)\sum_{i,j}n_{i,j}^h\nonumber\\
&-t_1\sum_{i,j}\left(h_{i+1,j}^{\dagger}h_{i,j}+h_{i,j+1}^{\dagger}h_{i,j}+{\rm H.c.}\right)\nonumber\\
&+V_1\sum_{i,j}\left(n_{i,j}^h n_{i+1,j}^h +n_{i,j}^h n_{i,j+1}^h \right)\nonumber\\
&-t_2\sum_{i,j}\left(h_{i+1,j+1}^{\dagger}(1-n_{i+1,j}^h-n_{i,j+1}^h) h_{i,j} \right . \nonumber\\
&~~~~~~~~~~~~ \left . +h_{i-1,j+1}^{\dagger}(1-n_{i-1,j}^h - n_{i,j+1}^h) h_{i,j}+{\rm H.c.}\right)\nonumber\\
&+V_2\sum_{i,j}\left(n_{i,j}^h n_{i+1,j+1}^h +n_{i,j}^h n_{i-1,j+1}^h \right)\nonumber\\
&-t_3\sum_{i,j}\left(h_{i+2,j}^{\dagger}(1-2n_{i+1,j}^h) h_{i,j} \right . \nonumber\\
&~~~~~~~~~~~~ \left . +h_{i,j+2}^{\dagger}(1-2n_{i,j+1}^h) h_{i,j}+{\rm H.c.}\right)\nonumber\\
&+V_3\sum_{i,j}\left(n_{i,j}^h n_{i+2,j}^h +n_{i,j}^h n_{i,j+2}^h \right),
\label{H_final2}
\end{align}
 with $\gamma=0=\eta$ and
with $t$ and $g$ being replaced by $t^\prime$ and $g^\prime$, respectively.
Since, the interaction terms  for the CJT Hamiltonian of LSNO are the same as those for
the $t_1-t_2-t_3-V_1-V_2-V_3$ Hamiltonian
in Eq. (\ref{H_final}), in LSNO also we expect to get the same charge-ordered phases obtained for the
$t_1-t_2-t_3-V_1-V_2-V_3$ model. Thus.  at hole doping $1/2$ and $1/3$
(i.e., at $x=1/2$ and $x=1/3$ in $\rm{La_{2-x}Sr_x NiO_4}$),
 we will realize checkerboard solid and diagonal stripes, respectively, which match exactly with the charge ordering 
 obtained for LSNO experimentally.

Now, if we add one extra hole to the system at one-third hole doping (i.e., at $x=1/3$),
then the extra hole will reside in the region between two diagonal stripes. This extra hole 
can hop anywhere in the region between the stripes without changing the potential energy of the system. 
Thus, the carriers for the hole doping $x>1/3$ are holes. On the other hand,  adding one electron to the 
striped phase so that $x<1/3$ will result in the extra electron  occupying any one of the sites along the stripes;
this extra electron is free to hop along the stripes without altering the potential energy of the system.
This means that electrons are the carriers for the doping $x<1/3$. Therefore, based on our model
we can explain the hole or electron doping (into the charge-ordered Mott insulator $\rm{La_{5/3}Sr_{1/3} NiO_4}$)
that was reported by T. Katsufuji {\it et al.} \cite{Hall_seebeck}

One can obviously ask how a system of HCBs can reproduce some experimental results of a system of electrons.
The reason behind the charge orderings at hole-doping values $1/2$ and $1/3$ is repulsion; 
hopping does not play any role in the ordering. Hence, for these two CDWs,
it does not matter whether the carriers
of the system are HCBs or electrons. Close to one-third doping,
only single carrier physics plays a role; consequently, particle-hole asymmetry is captured. 
Next, it is important to note that  CJT interaction is needed to generate
NNN and NNNN repulsions $V_2$ and  $V_3$ which in turn are needed to explain
diagonal stripes.
Thus, we see that our work resolves the controversy whether
 cooperative Jahn-Teller distortions can explain the observed diagonal-stripe charge order at one-third
 doping in LSNO  \cite{dagotto,hatsugai,oles}.
 {Lastly, it should also be pointed out that, although experimentally \cite{tokura2}
 insulating behavior
 is observed in LSNO for $x \lesssim 0.9$, theoretically we expect metallic nature; we believe,
 this discrepancy is due to localization effects produced by disorder in real LSNO systems.}

\section{Conclusions and open problems}\label{Conclusion}
To conclude, we investigated a 2D system of HCBs, modulated by the cooperative breathing mode, which is important in real
materials such as  $\rm{BaBiO_3}$ and nickelates as well as in artificial cold-atom systems. Using a 
duality treatment, we obtained the effective Hamiltonian and generated the phase diagram employing the SSE technique.

In the phase diagram
displayed in Fig. \ref{phase_diagram},  a first-order transition occurs from a superfluid to a checkerboard
solid at filling-fraction 1/2  and from a superfluid to a diagonal striped
solid at filling 1/3. We interpreted the nature of the transition by
invoking Landau's explanation. It would be interesting to verify whether
in other unfrustrated lattices, such as the checkerboard lattice,  a discontinuous superfluid-solid transition
is manifested at commensurate fillings such as 1/4 \cite{wessel4}.
{Furthermore, at a fixed interaction strength,  our $t_1-t_2-t_3-V_1-V_2-V_3$ model realizes
only continuous transitions (i.e., superfluid-supersolid and supersolid-solid transitions)
as density is varied. Contrastingly,  the $t_1-t_2-t_3-V_1$ model (pertaining to the strong-coupling case
of  the Holstein model) 
manifests a discontinuous superfluid-supersolid
transition when density is varied \cite{kar}. Thus, more insight is needed to
identify which class of models yield what type of superfluid-supersolid transition.}  

We have identified the $t_1-V_1-V_2-V_3$ model as the minimum model for obtaining a diagonal striped supersolid on
a square lattice.
It would be exciting to realize this system  in a cold-atom system, thereby adding to the
understanding of lattice supersolidity generated by long-range interactions \cite{expt_supersolid}.

The asymmetry of the supersolid phase about  a commensurate filling, such as  one third in our case
and one fourth in the case of  Ref. \onlinecite{boninsegni1}, in a square lattice occurs 
possibly because particle-hole symmetry
is not respected by the Hamiltonian about these fillings. It would be worthwhile to
study the nature of such asymmetry in other lattices such as honeycomb, checkerboard, etc.

We have explained the charge ordering in ${\rm La}_{2-x}{\rm Sr}_x{\rm NiO_4}$ at hole-doping
$x=1/2$ and $1/3$ by considering cooperative Jahn-Teller effect. However, studies
involving CJT effect
are needed  at
dopings away from these fillings and  particularly in the region   $x > 1/2$ where holes are doped into the
$d_{z^2}$ orbitals. Also of interest would the explanation for the
metal-insulator transition observed at $x \sim 0.9$
\cite{tokura2}.

In a different  but related system  $\rm{La_{2-x}Sr_xCoO_4}$,  CDWs similar to those in LSNO
are observed.
At half doping, there is a signature of checkerboard charge ordering with alternate
$\rm Co^{2+}$ and $\rm Co^{3+}$ ions (below $T_{\rm CO}\approx 750$K) \cite{cobaltate_ckh_brd}. 
On the other hand, at the doping $x=1/3$,  the holes form a diagonal-stripe pattern similar to the 
stripes in LSNO at a transition temperature well above the room temperature\cite{cobaltate_stripe1,cobaltate_stripe2,cobaltate_stripe3,cobaltate_stripe4}. 
Furthermore, the presence of substantial disorder in these diagonal stripes has been confirmed by the experiment \cite{cobaltate_stripe4} done by A. T. Boothroyd {\it et al.} 
The electronic configuration of cobalt is ${\rm [Ar] 3d^7 4s^2}$. In $\rm{La_{2-x}Sr_xCoO_4}$, cobalt shows two
different oxidation states: $\rm Co^{2+}$ and $\rm Co^{3+}$. The $\rm Co^{3+}$ ions are found to have 
the low-spin ground state (${\rm S=0}$) \cite{cobaltate_oxidation} with the electronic configuration ${\rm [Ar] 3d^6}$. 
In this case, all the six $d$ electrons occupy the $t_{2g}$ orbitals and both the $e_g$ orbitals are empty. 
Therefore, $\rm Co^{3+}$ ions do not cause any Jahn-Teller distortion in the system. On the other hand, 
in the case of $\rm Co^{2+}$ ions, the electrons are in the high-spin ground state (${\rm S=3/2}$)
with the electronic configuration ${\rm [Ar] 3d^7}$. This state consists of five electrons 
in the $t_{2g}$ orbitals and two in the $e_g$ orbitals. Two out of the three $t_{2g}$ orbitals 
are completely filled with four electrons, whereas the remaining orbital contains a single electron.
Since both the  $e_g$ orbitals are occupied by one electron each, JT distortion comes into play 
due to the singly occupied $t_{2g}$ orbital only. Owing to the fact that the JT distortion 
arising from $t_{2g}$ electrons is weaker than the one arising from $e_g$ electrons, 
it needs to be examined whether this can explain the disorder in
the stripe pattern in $\rm{La_{2-x}Sr_xCoO_4}$.

\section{Acknowledgements}
The computing resources of the Condensed Matter Physics Division (Saha Institute of Nuclear Physics)
have been used extensively. Valuable discussions with R. Pankaj are acknowledged. A.G. would especially like to thank S. Kar, G. Majumdar and M. Sarkar for useful discussions regarding SSE. S.Y. thanks P. B. Littlewood for stimulating
discussions and Cavendish lab for hospitality during the initial stages of this work.

\appendix

\section{Nearest-neighbor repulsion}
The second order perturbation term is given by
\begin{align}
H^{(2)}=-\sum_m\sum_{i,j,k,l}&\frac{\langle 0|_{ph} H_{1i,j} |m\rangle_{ph}\langle m|_{ph} H_{1k,l} |0\rangle_{ph}}{E_0^{ph}-E_m^{ph}}\nonumber\\
=-t_1^2 \sum_m\sum_{i,j,k,l}\frac{1}{\Delta E_m^{ph}}&\Big[\Big( d_{i+1,j}^\dagger d_{i,j}\langle 0|_{ph}(\tau_{-x}^{ij}-1)|m\rangle_{ph}\nonumber\\
+&d_{i,j}^\dagger d_{i+1,j}\langle 0|_{ph}(\tau_{+x}^{ij}-1)|m\rangle_{ph}\nonumber\\
+&d_{i,j+1}^\dagger d_{i,j}\langle 0|_{ph}(\tau_{-y}^{ij}-1)|m\rangle_{ph}\nonumber\\
+&d_{i,j}^\dagger d_{i,j+1}\langle 0|_{ph}(\tau_{+y}^{ij}-1)|m\rangle_{ph}\Big)\nonumber\\
\times\Big(&d_{k+1,l}^\dagger d_{k,l}\langle m|_{ph}({\tau_{+x}^{kl}}^\dagger-1)|0\rangle_{ph}\nonumber\\
+&d_{k,l}^\dagger d_{k+1,l}\langle m|_{ph}({\tau_{-x}^{kl}}^\dagger-1))|0\rangle_{ph}\nonumber\\
+&d_{k,l+1}^\dagger d_{k,l}\langle m|_{ph}({\tau_{+y}^{kl}}^\dagger-1))|0\rangle_{ph}\nonumber\\
+&d_{k,l}^\dagger d_{k,l+1}\langle m|_{ph}({\tau_{-y}^{kl}}^\dagger-1))|0\rangle_{ph}\Big)\Big],
\label{H2_app}
\end{align}
where $t_1=te^{-(E_p+V_p)/\omega_0}$ and $\Delta E_m^{ph}=E_0^{ph}-E_m^{ph}$.

As already mentioned in Sec. \ref{subsec_NN_repul}, the NN repulsion results from a process where a particle hops to its neighboring site and returns back, which in 2D consists of two terms: $\sum\limits_{i,j}\left[n_{i,j}(1-n_{i+1,j})+n_{i+1,j}(1-n_{i,j})\right]$ and $\sum\limits_{i,j}\left[n_{i,j}(1-n_{i,j+1})+n_{i,j+1}(1-n_{i,j})\right]$.

Since, $\sum\limits_{i,j}n_{i,j}(1-n_{i+1,j})=\sum\limits_{i,j}n_{i+1,j}(1-n_{i,j})$ and $\sum\limits_{i,j}n_{i,j}(1-n_{i,j+1})=\sum\limits_{i,j}n_{i,j+1}(1-n_{i,j})$, so the process is effectively given by 
{ $\sum\limits_{i,j}\left[n_{i,j}(1-n_{i+1,j})+n_{i,j}(1-n_{i,j+1})\right]$} with the coefficient being twice.

Now, we can rewrite the term $\sum\limits_{i,j}n_{i,j}(1-n_{i+1,j})$ as 
\begin{align}
\sum\limits_{i,j}d^\dagger_{i,j}d_{i,j}(1-d^\dagger_{i+1,j}d_{i+1,j})&=\sum\limits_{i,j}d^\dagger_{i,j}d_{i,j}d_{i+1,j}d^\dagger_{i+1,j}\nonumber\\
&=\sum\limits_{i,j}d^\dagger_{i,j}d_{i+1,j}d^\dagger_{i+1,j}d_{i,j}.\nonumber
\end{align}
Looking at the expression of $H^{(2)}$, one can figure out that the above term comes from the multiplication of the terms $d^\dagger_{i,j}d_{i+1,j}$ and $d^\dagger_{k+1,l}d_{k,l}$ for $k=i$ and $l=j$. So, the coefficient of this
term is given by
\begin{align}
 t_1^2\sum_m\frac{\langle 0|_{ph}(\tau_{+x}^{ij}-1)|m\rangle_{ph}\langle m|_{ph}({\tau_{+x}^{ij}}^\dagger-1)|0\rangle_{ph}}{\Delta E_m^{ph}} ,
\end{align}
where
\begin{align}
&\tau^{ij}_{+ x}= \exp \Big[ g(2a_{i,j}-a_{i-1,j}-a_{i+1,j})\nonumber\\
&+g(b_{i+1,j-1}+b_{i,j}-b_{i,j-1}-b_{i+1,j})+ \gamma g(c_{i,j}-c_{i+1,j})\Big] ;\nonumber
\end{align}
consequently, the coefficient simplifies exactly to be $\frac{t_1^2}{\omega_0}G_9(4,1,1,1,1,1,1,\gamma^2,\gamma^2)$.
Now, the general form  $G_n(\alpha_1,\alpha_2,\cdots,\alpha_n)$ can be expressed as
\begin{align}
\!\!\!\!
G_n(\alpha_1,\alpha_2,\cdots,\alpha_n)
 =&\sum_{m_1,m_2,...,m_n}^{\prime}\frac{(\alpha_1 g^2)^{m_1}\cdots(\alpha_n g^2)^{m_n}}
 {m_1!\cdots m_n!(m_1+\cdots+m_n)} ,\nonumber
\end{align}
where $m_i = 0,1,2,....,\infty$ and the prime in $\sum^{\prime}$ 
implies the case $m_1=m_2=...=m_n=0$ is excluded
from the summation.
It is important to note that for large values of $g^2$, $G_n$ can be approximately expressed as
\begin{align}
G_n(\alpha_1,\alpha_2,\cdots,\alpha_n) \approx \frac{\exp\left(\sum\limits_{i=1}^n\alpha_i g^2\right)}{\sum\limits_{i=1}^n\alpha_i g^2}.\label{G_napprox}
\end{align}
{
Then, the NN repulsion is given by
\begin{align}
-V_z\sum\limits_{i,j}[n_{i,j}(1-n_{i+1,j})+n_{i,j}(1-n_{i,j+1})], \label{NN_repul}
\end{align}
where
\begin{eqnarray}
V_z&=&\frac{2t^2e^{-2(E_p+V_p)/\omega_0}}{\omega_0}G_9(4,1,1,1,1,1,1,\gamma^2,\gamma^2) \nonumber \\
&\approx& \frac{2t^2}{2E_p+2V_p} .
\end{eqnarray}
}

Now, in arriving at Eq. (\ref{NN_repul}), we did not take into account the occupancy
of the neighbors of the intermediate site. For example, when the particle hops from site
$(i,j)$ to NN site $(i+1,j)$ and back, we have not considered the occupancy  
of the sites $(i + 2, j)$, $(i + 1, j + 1)$ and $(i + 1, j - 1)$, which are the neighboring sites of the 
intermediate site $(i + 1, j)$ (as can be seen from Fig. \ref{2d_CBM_fig}).
We will consider this occupancy  in the next Appendix. 

\section{NNN repulsion 
and NNNN repulsion 
}
In this appendix we first outline the procedure of calculating the coefficient of next-nearest-neighbor (NNN) repulsion which occurs along the diagonals.
\begin{figure}[t]
\includegraphics[width=0.5\linewidth]{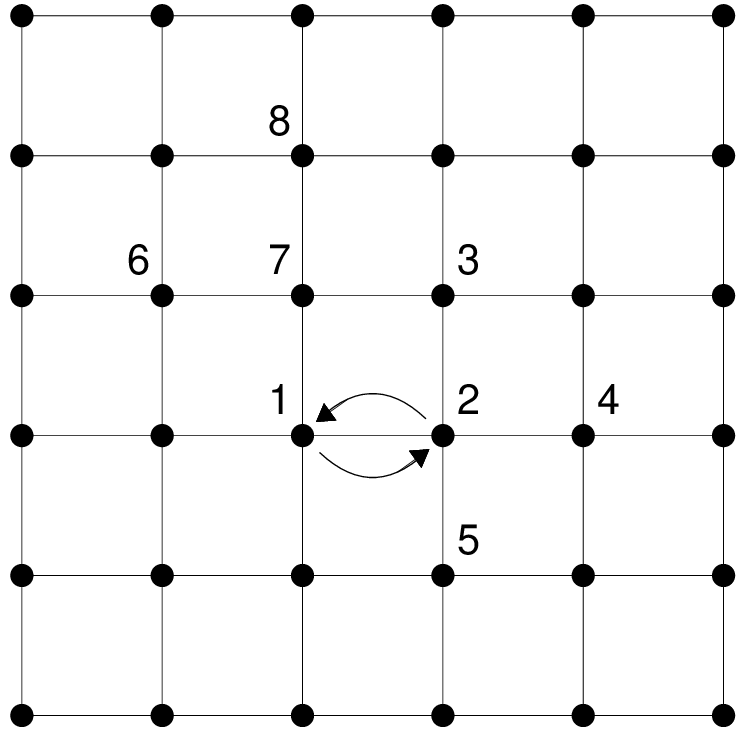}
\caption{Pictorial description of the process where a particle at site 1 hops to site 2 and comes back.}\label{pictorial_1}
\end{figure}
Consider the case where a particle hops to its neighboring site and returns back yielding the term $\propto\sum\limits_{<i,j>}n_i(1-n_j)$ with $<i,j>$ indicating nearest-neighbor (NN) pairs of sites. 
In this process we have to take into account the occupancy of the neighboring sites of the intermediate site $j$.
For example, in  Fig. \ref{pictorial_1}, if a particle at site 1 hops to site 2 and comes back, 
then the coefficient of this process depends on the occupancy of the sites 3, 4, and 5.
If all the three sites are empty, then this term can be expressed as $-V_z n_1(1-n_3)(1-n_4)(1-n_5)$ where $V_z\approx\frac{2t^2}{2E_p+2V_p}$; here, 
we have omitted the term $(1-n_2)$ because the  possibility of NN occupancy (for particle at site 1) is
already excluded from the process
 due to the large value of NN repulsion $2V_p$. 
 Due to numerical difficulties in our simulations using SSE,
 we need to simplify the four-operator term into a two-operator one by applying
 mean field to the remaining two operators. One can easily see that this mean-field procedure 
 leaves us with a term which represents NNN repulsion 
(which acts along the diagonals) or NNNN repulsion (which acts along the axes).

We will now  calculate the NNN repulsion coefficient which
pertains to the diagonals of 
the square lattice in  Fig. \ref{pictorial_1}. To this end, we consider all the possible processes yielding the operator $n_1n_3$  
and add all the corresponding  terms to evaluate its coefficient. The following are the relevant
cases.\\
Case 1 :  NNN interaction, 
when all the three neighboring sites of the intermediate site are unoccupied,
involves the following.\\ 
(i) The contribution of particle hopping from site 1 to site 2 and coming back:
 \begin{align}
 &=-\frac{2t^2}{(2E_p+2V_p)}n_1(1-n_3)(1-n_4)(1-n_5)\nonumber\\
 &\approx-\frac{2t^2}{(2E_p+2V_p)}n_1(1-n_3)\langle 1-n_4\rangle\langle 1-n_5\rangle \nonumber \\
 &\approx-\frac{2t^2}{(2E_p+2V_p)}n_1(1-n_3)\left (\frac{1}{2}-m \right )^2 ,
 \end{align}
 where  $\langle ... \rangle$ implies mean value and $\langle 1-n_4\rangle = \langle 1-n_5\rangle
 = (\frac{1}{2}-m) $ with $m$ being the magnetization of the system.\\
 \\
 (ii) The contribution of particle hopping from site 1 to site 7 and returning back involves a similar situation
 as (i) and is given as:
 \begin{align}
 &\approx-\frac{2t^2}{(2E_p+2V_p)}n_1(1-n_3)\langle 1-n_6\rangle\langle 1-n_8\rangle \nonumber \\
 &\approx-\frac{2t^2}{(2E_p+2V_p)}n_1(1-n_3)\left (\frac{1}{2}-m \right )^2 .
 \end{align}
 (iii) The contribution of particle hopping from site 3 to site 2 and coming back:
 \begin{align}
 &\approx -\frac{2t^2}{(2E_p+2V_p)}n_3(1-n_1)\langle 1-n_4\rangle\langle 1-n_5\rangle \nonumber \\
&\approx-\frac{2t^2}{(2E_p+2V_p)}n_3(1-n_1)\left (\frac{1}{2}-m \right )^2 .
\label{b3}
 \end{align}
 (iv) The particle hopping from site 3 to site 7 and returning back is similar
 to (iii) and yields the same expression as Eq. (\ref{b3}).
  \\
 (v) The contribution of particle hopping from site 4 to site 2 and coming back:
 \begin{align}
 &\approx -\frac{2t^2}{(2E_p+2V_p)}(1-n_3)(1-n_1)\langle n_4\rangle\langle 1-n_5\rangle \nonumber \\
 &\approx -\frac{2t^2}{(2E_p+2V_p)}(1-n_3)(1-n_1) \left ( \frac{1}{4} - m^2 \right )^2 .
 \label{b5}
 \end{align}
  (vi) The contribution of particle hopping from site 5 to site 2 and coming back
  is similar to (v) and is given by Eq.(\ref{b5}).
  \\
  (vii) The particle hopping from site 6 to site 7 and returning back is also similar to (v)
  and the contribution is again given by Eq.(\ref{b5}).
 \\ 
 (viii) The contribution of particle hopping from site 8 to site 7 and coming back
 is also similar to (v) and  hence is given by Eq.(\ref{b5}).
\\
Adding all the contributions for case 1,
we get the coefficient of $n_1n_3$ to be
\begin{align}
 \frac{2t^2}{(2E_p+2V_p)}\left[4\left(\frac{1}{2}-m\right)^2-4\left(\frac{1}{4}-m^2\right)\right].\label{occupied_0}
\end{align}

Case 2: 
We consider contribution
to NNN interaction when, among the three sites that are NN to the intermediate site, one of them  
is occupied and the other two are empty. Thus, compared to case 1, there is 
an extra repulsion term $2V_p$ in the denominator of the coefficient.
 Then, NNN interaction 
involves the following.\\
(i) The particle hops from site 1 to site 2 and comes back. Any one of the three neighboring sites of site 2, 
i.e., 3, 4, or 5, is occupied; then, the contribution is:
\begin{align}
 &\approx -\frac{2t^2}{(2E_p+4V_p)}\Big[n_1n_3\langle 1-n_4\rangle\langle 1-n_5\rangle\nonumber\\
 &+n_1(1-n_3)\langle n_4\rangle\langle 1-n_5\rangle+n_1(1-n_3)\langle 1-n_4\rangle \langle n_5\rangle\Big]\nonumber\\
 &\approx-\frac{2t^2}{(2E_p+4V_p)}\Bigg[n_1n_3\left(\frac{1}{2}-m\right)^2\nonumber\\
 &\qquad\qquad\qquad\qquad+2n_1(1-n_3)\left(\frac{1}{4}-m^2\right)\Bigg].
  \label{b_2_1}
\end{align}
\\
(ii) The particle hops from site 1 to site 7 and comes back. The situation is similar to (i) and hence
the contribution is given by Eq. (\ref{b_2_1}).
\\
(iii) The particle hops from site 3 to site 2 and comes back. The resulting contribution is
\begin{align}
&\approx -\frac{2t^2}{(2E_p+4V_p)}\Big[n_3 n_1\langle 1-n_4\rangle\langle 1-n_5\rangle\nonumber\\
 &~~+n_3(1-n_1)\langle n_5\rangle\langle 1-n_4\rangle+n_3(1-n_1)\langle n_4\rangle\langle 1-n_5\rangle\Big]\nonumber\\
 & \approx
-\frac{2t^2}{(2E_p+4V_p)}\Bigg[n_1n_3\left(\frac{1}{2}-m\right)^2\nonumber\\
 &\qquad\qquad\qquad\qquad+2n_3(1-n_1)\left(\frac{1}{4}-m^2\right)\Bigg].
  \label{b_2_3}
\end{align}
\\
(iv) The particle hops from site 3 to site 7 and returns back. Since the situation is similar to (iii),
the contribution is expressed by Eq. (\ref{b_2_3}).
\\
(v) The particle hops from site 4 to site 2 and comes back. The contribution is
\begin{align}
&\approx -\frac{2t^2}{(2E_p+4V_p)}\Big[\langle n_4\rangle (1-n_1)n_3\langle 1-n_5\rangle\nonumber\\
 &+\langle n_4\rangle n_1(1-n_3)\langle 1-n_5\rangle+\langle n_4\rangle (1-n_1)(1-n_3)\langle n_5\rangle\Big]\nonumber\\
  &\approx -\frac{2t^2}{(2E_p+4V_p)}\Bigg[n_1(1-n_3)\left(\frac{1}{4}-m^2\right)\nonumber\\
 &+n_3(1-n_1)\left(\frac{1}{4}-m^2\right)+(1-n_1)(1-n_3)\left(\frac{1}{2}+m\right)^2\Bigg].
 \label{b_2_5}
 \end{align}
\\
(vi) The particle hops from site 5 to site 2 and comes back. The situation being similar to
(v) leads to the contribution being given by Eq. (\ref{b_2_5}).
\\
(vii) The particle hops from site 6 to site 7 and comes back; this circumstance is also similar to (v)
and hence contribution same as in Eq. (\ref{b_2_5}).
\\
(viii) The particle hops from site 8 to site 7 and comes back. Here too the contribution
is given by Eq. (\ref{b_2_5}) since the circumstance is again similar to (v).
\\
Therefore, for case 2, the sum total of the above contributions yields the coefficient of $n_1n_3$ to be
\begin{align}
\frac{2t^2}{(2E_p+4V_p)}\Bigg[16\left(\frac{1}{4}-m^2\right)&-4\left(\frac{1}{2}-m\right)^2\nonumber\\
&-4\left(\frac{1}{2}+m\right)^2\Bigg]. \label{occupied_1}
\end{align}

Case 3: Contribution to NNN interaction when the intermediate site
has any two of the three NN sites  occupied with the other being empty. 
Then, compared to case 2, the coefficient has an extra repulsion term $2V_p$ in the denominator;
consequently, NNN interaction 
involves the following.\\
(i) The particle hops from site 1 to site 2 and comes back; the resulting contribution is:
\begin{align}
  &\approx 
 -\frac{2t^2}{(2E_p+6V_p)}\Big[n_1n_3\langle n_4\rangle\langle 1-n_5\rangle\nonumber\\
 &\qquad\qquad+n_1n_3\langle 1-n_4\rangle\langle n_5\rangle+n_1(1-n_3)\langle n_4\rangle \langle n_5\rangle\Big]\nonumber\\
 &\approx
 -\frac{2t^2}{(2E_p+6V_p)}\Bigg[2n_1n_3\left(\frac{1}{4}-m^2\right)\nonumber\\
 &\qquad\qquad\qquad\qquad+n_1(1-n_3)\left(\frac{1}{2}+m\right)^2\Bigg].
  \label{b_3_1}
\end{align}
\\
(ii) The particle hops from site 1 to site 7 and comes back. This situation is similar
to (i) with the contribution being expressed by Eq. (\ref{b_3_1}).
\\
(iii) The particle hops from site 3 to site 2 and returns; the ensuing contribution is:
\begin{align}
 &\approx -\frac{2t^2}{(2E_p+6V_p)}\Big[n_3 n_1\langle 1-n_4\rangle\langle n_5\rangle\nonumber\\
 &\qquad\qquad+n_3 n_1\langle n_4\rangle\langle 1-n_5\rangle+n_3(1-n_1)\langle n_4\rangle\langle n_5\rangle\Big]\nonumber\\
  &\approx -\frac{2t^2}{(2E_p+6V_p)}\Bigg[2n_1n_3\left(\frac{1}{4}-m^2\right)\nonumber\\
 &\qquad\qquad\qquad\qquad+n_3(1-n_1)\left(\frac{1}{2}+m\right)^2\Bigg].
 \label{b_3_2}
 \end{align}
\\
(iv) The particle hops from site 3 to site 7 and comes back. The situation is similar to (iii)
with the contribution being given by Eq. (\ref{b_3_2}).
\\
(v) The particle hops from site 4 to site 2 and returns. This produces the contribution:
\begin{align}
 &\approx -\frac{2t^2}{(2E_p+6V_p)}\Big[\langle n_4\rangle n_1 n_3\langle 1-n_5\rangle\nonumber\\
 &\qquad\qquad+\langle n_4\rangle (1-n_1)n_3\langle n_5\rangle+\langle n_4\rangle n_1(1-n_3)\langle n_5\rangle\Big]\nonumber\\
  &\approx -\frac{2t^2}{(2E_p+6V_p)}\Bigg[n_1n_3\left(\frac{1}{4}-m^2\right)\nonumber\\
 &\quad+n_1(1-n_3)\left(\frac{1}{2}+m\right)^2+n_3(1-n_1)\left(\frac{1}{2}+m\right)^2\Bigg].
 \label{b_3_3}
 \end{align}
\\
(vi) The particle hops from site 5 to site 2 and comes back. The circumstance, being similar to (v),
yields the contribution expressed in Eq. (\ref{b_3_3}).
\\
(vii) The particle hops from site 6 to site 7 and comes back. The situation is also similar to (v)
with the contribution being also given by Eq. (\ref{b_3_3}).
\\
(viii) The particle hops from site 8 to site 7 and returns. Again the situation is similar
to (v) with the contribution being again given by Eq. (\ref{b_3_3}).
\\
Therefore, on adding all the various contributions for case 3, we get the coefficient of $n_1n_3$ to be
\begin{align}
\frac{2t^2}{(2E_p+6V_p)}\Bigg[12\left(\frac{1}{2}+m\right)^2-12\left(\frac{1}{4}-m^2\right)\Bigg]. \label{occupied_2}
\end{align}

Case 4: Contribution to NNN interaction when all of the three neighboring sites of the intermediate
site are occupied. Here, compared to case 3, the coefficient has an extra repulsion term $2V_p$ in the denominator.
Then,  NNN interaction involves the following.\\
(i) The particle hops from site 1 to site 2 and comes back. Consequently,
the contribution is 
\begin{align}
&\approx
-\frac{2t^2}{(2E_p+8V_p)} n_1 n_3 \langle n_4\rangle\langle n_5\rangle\nonumber\\
&\approx
-\frac{2t^2}{(2E_p+8V_p)} n_1 n_3\left(\frac{1}{2}+m\right)^2.
\label{b_4_1}
\end{align}
\\
For all the following also the contribution is expressed by Eq. (\ref{b_4_1}) because
the situation is similar to (i).\\ 
(ii) The particle hops from site 1 to site 7 and comes back.\\
(iii) The particle hops from site 3 to site 2 and returns. \\
(iv) The particle hops from site 3 to site 7 and comes back. \\
(v) The particle hops from site 4 to site 2 and comes back. \\
(vi) The particle hops from site 5 to site 2 and returns. \\
(vii) The particle hops from site 6 to site 7 and comes back. \\
(viii) The particle hops from site 8 to site 7 and returns.\\
Therefore, for case 4, the coefficient of $n_1 n_3$ is given by
\begin{align}
-\frac{2t^2}{(2E_p+8V_p)}\times 8\left(\frac{1}{2}+m\right)^2. \label{occupied_3}
\end{align}

Combining Eqs. (\ref{occupied_0}), (\ref{occupied_1}), (\ref{occupied_2}) and (\ref{occupied_3}), we finally get the coefficient of NNN repulsion (which acts along the diagonals) to be
\begin{align}
V_2&=2t^2\Bigg[\left(\frac{1}{2}-m\right)^2\frac{2V_p}{(E_p+V_p)(E_p+2V_p)}\nonumber\\
&\qquad+\left(\frac{1}{4}-m^2\right)\frac{4E_pV_p}{(E_p+V_p)(E_p+2V_p)(E_p+3V_p)}\nonumber\\
&\qquad+\left(\frac{1}{2}+m\right)^2\frac{2E_pV_p}{(E_p+2V_p)(E_p+3V_p)(E_p+4V_p)}\Bigg].
\end{align}

To calculate the NNNN repulsion along the x-axis (y-axis), we have to consider all the processes from which a 
term $n_1 n_4$ ($n_1 n_8$) can appear. Adding all those terms, we can see that the coefficient of NNNN repulsion
is just half of the coefficient of NNN repulsion. The reason for this is that the relevant
contributions are from only 
half of the eight situations considered in each of the above four occupancy cases (i.e, the
four cases involving different number of occupied neighbors for the intermediate site).

\section{NNN hopping 
and NNNN hopping }
\begin{figure}[t]
\includegraphics[width=0.5\linewidth]{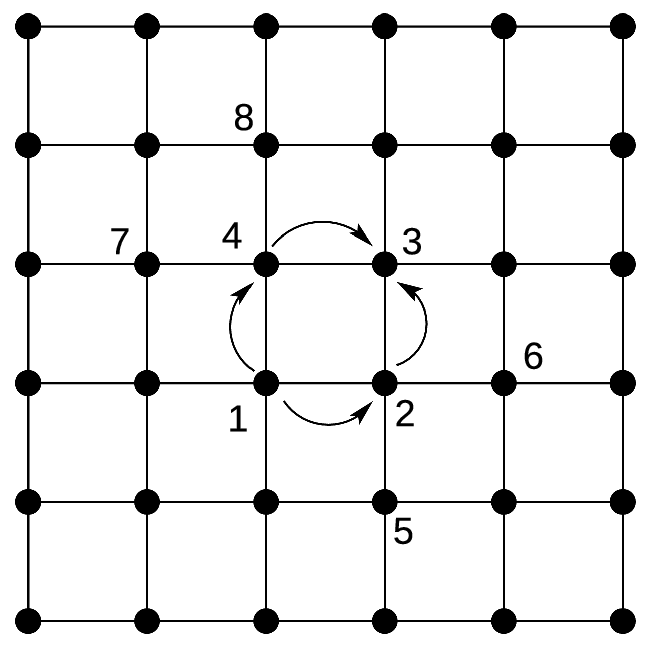}
\caption{Pictorial depiction of the process where a particle at site 1 hops to site 3 which is its NNN site along diagonal. The two possible paths for this process are indicated: hopping to site 3 via site 2 and site 4. }\label{pictorial_2}
\end{figure}

There are two possible hopping paths for a particle  to arrive at a NNN site along the diagonals of the square lattice. 
For example, in Fig. \ref{pictorial_2}, consider a particle hopping from site 1 to site 3. It can either hop to site 2
first and then to site 3 or it can hop to site 4 followed by a hop to site 3. Now, the coefficient of this process gets
modified by the occupancy of the neighboring sites of the intermediate site. Without taking into account this effect,
the process along any one path [on using Eq. (\ref{H2_app})]
is given exactly by
\begin{eqnarray*}
-\frac{t^2e^{-2(E_p+V_p)/\omega_0}}{\omega_0}G_5(2,2,1,1,\gamma^2)
\sum\limits_{<<i,j>>} (d_i^\dagger d_j+ {\rm H.c.}) ,
\end{eqnarray*}
where $<<i,j>>$ denotes NNN pairs of sites along the diagonals.
For large values of $g^2$, we have the following simplification for the coefficient in the above expression: 
\begin{eqnarray*}
\frac{t^2e^{-2(E_p+V_p)/\omega_0}}{\omega_0}G_5(2,2,1,1,\gamma^2)
\approx \frac{t^2 e^{-E_p/\omega_0}}{E_p+2V_p} .
\end{eqnarray*}

\noindent Path 1: The particle hops from site 1 to site 3 via site 2. The coefficient of this process depends on the occupancy of the sites 5 and 6 which are the two neighboring sites of the intermediate site 2.
 
Case 1: Contribution to NNN hopping when both the neighboring sites are empty:
\begin{align}
&-\frac{t^2 e^{-E_p/\omega_0}}{E_p+2V_p}d_3^\dagger d_1 (1-n_5)(1-n_6)\nonumber\\
&\approx
 -\frac{t^2 e^{-E_p/\omega_0}}{E_p+2V_p}d_3^\dagger d_1 \langle 1-n_5\rangle\langle 1-n_6\rangle\nonumber\\
& \approx -\frac{t^2 e^{-E_p/\omega_0}}{E_p+2V_p}\left(\frac{1}{2}-m\right)^2 d_3^\dagger d_1.
\label{c1}
\end{align}

Case 2: Contribution when any one of the neighboring sites is occupied (giving an extra repulsion $2V_p$ in the denominator) and the other site is empty:
\begin{align}
&\approx -\frac{t^2 e^{-E_p/\omega_0}}{E_p+4V_p}d_3^\dagger d_1 \left[\langle n_5\rangle\langle 1-n_6\rangle+\langle 1-n_5\rangle \langle n_6\rangle\right]\nonumber\\
& \approx -\frac{2t^2 e^{-E_p/\omega_0}}{E_p+4V_p}\left(\frac{1}{4}-m^2\right) d_3^\dagger d_1.
\label{c2}
\end{align}

Case 3: Contribution when both the NN sites are occupied:
\begin{align}
\approx& -\frac{t^2 e^{-E_p/\omega_0}}{E_p+6V_p}d_3^\dagger d_1 \langle n_5\rangle\langle n_6\rangle\nonumber\\
&\approx
 -\frac{t^2 e^{-E_p/\omega_0}}{E_p+6V_p}\left(\frac{1}{2}+m\right)^2 d_3^\dagger d_1.
 \label{c3}
\end{align}

Therefore, for path 1, we get the coefficient of $d_3^\dagger d_1$ to be
\begin{align}
-&t^2 e^{-E_p/\omega_0}\Bigg[\left(\frac{1}{2}-m\right)^2\frac{1}{E_p+2V_p}\nonumber\\
&+\left(\frac{1}{4}-m^2\right)\frac{2}{E_p+4V_p}+\left(\frac{1}{2}+m\right)^2\frac{1}{E_p+6V_p}\Bigg].
\label{process_1}
\end{align}

\noindent Path 2: The particle hops from site 1 to site 4 first and then to site 3. The coefficient of this process gets modified depending on whether the sites 7 and 8 (NN to the intermediate site 4) are occupied or not.

Case 1 : Contribution when both the neighboring sites are empty. This situation is similar
to case 1 of path 1; hence, the contribution is given by Eq. (\ref{c1}).

Case 2 : Contribution when any one of the neighboring sites is occupied and the other one is empty. This
is similar to case 2 of path 1; consequently, the contribution is expressed by Eq. (\ref{c2}).

Case 3 : Contribution when both the NN sites are occupied. This circumstance is similar to case 3 of path 1;
thus, the contribution is given by Eq. (\ref{c3}).

Thus we see that path 2 yields the same coefficient [given by Eq. (\ref{process_1})]
for $d_3^\dagger d_1$ as path 1.
Combining 
the contributions from both the  paths, a particle hopping to its NNN along diagonals can be expressed as
$-t_2\sum\limits_{<<i,j>>} (d_i^\dagger d_j+ {\rm H.c.})$, where the coefficient $t_2$ is given by
\begin{align}
 t_2&=2t^2 e^{-E_p/\omega_0}\Bigg[\left(\frac{1}{2}-m\right)^2\frac{1}{E_p+2V_p}\nonumber\\
&+\left(\frac{1}{4}-m^2\right)\frac{2}{E_p+4V_p}+\left(\frac{1}{2}+m\right)^2\frac{1}{E_p+6V_p}\Bigg].
\end{align}

For the case of NNNN hopping (which occurs along the axes), there is only one possible path.
Hence, the relevant coefficient $t_3$ for NNNN hopping
is half of the coefficient for NNN hopping, i.e., $t_3 = \frac{t_2}{2}$.



\begin{thebibliography}{1}
\bibitem{min_cb}
In one dimension, it has been shown in Ref. \onlinecite{ghosh} that $t_2-V_1$ model
yields a cSS.
In 2D, it has been demonstrated in Ref. \onlinecite{kar} that $t_2-t_3-V_1$ model generates a cSS;
using similar logic, it can be shown that $t_2-V_1$ model also manifests cSS.
\end{thebibliography}
\end{document}